\documentclass[11pt]{article}
\usepackage{a4p}
\usepackage{epsfig}
\usepackage{array}
\usepackage{cite}
\usepackage{pennames}
\usepackage{rotating}
\newcommand{\PPEnum} {CERN-EP/2003-049}
\newcommand{\Date}      {11 July 2003}

\newcommand{\mybold} {\boldmath }

\def\lumic{29}
\def\lumid{73}
\def\lumie{74}
\def\lumif{37}
\def\lumig{80}  
\def\lumih{140}
\def\emnc{$191.59\pm0.04$}
\def\emnd{$195.53\pm0.04$}
\def\emne{$199.52\pm0.04$}
\def\emnf{$201.63\pm0.04$}
\def\emng{$204.80\pm0.05$}
\def\emnh{$206.53\pm0.05$}
 \def\xscc{1.29 \ ^{+ 0.47 }_{-0.40 }   \ ^{+ 0.12 }_{-0.09 }   }
 \def\xscd{1.13 \ ^{+ 0.26 }_{-0.24 }   \ ^{+ 0.08 }_{-0.07 }   }
 \def\xsce{1.05 \ ^{+ 0.25 }_{-0.22 }   \ ^{+ 0.07 }_{-0.06 }   }
 \def\xscf{0.79 \ ^{+ 0.35 }_{-0.29 }   \ ^{+ 0.08 }_{-0.05 }   }
 \def\xscg{1.07 \ ^{+ 0.27 }_{-0.24 }   \ ^{+ 0.08 }_{-0.07 }   }
 \def\xsch{0.97 \ ^{+ 0.19 }_{-0.18 }   \ ^{+ 0.06 }_{-0.05 }   }
 \newcommand{\lllllb} {a}
 \newcommand{\eennlb} {b}
 \newcommand{\mmnnlb} {c}
 \newcommand{\qqeelb} {d}
 \newcommand{\bbeelb} {e}
 \newcommand{\qqmmlb} {f}
 \newcommand{\bbmmlb} {g}
 \newcommand{\qqttlb} {h}
 \newcommand{\bbttlb} {i}
 
 \newcommand{\qqnnlb} {k}
 \newcommand{\bbnnlb} {l}
 \newcommand{\qqqqlb} {m}
 \newcommand{\qqxblb} {n}
 \newcommand{\qqbblb} {o}

\newcommand{\inmath}[1] {\ifmmode#1\else$#1$\fi}
\newcommand{\definmath}[2] {\def#1{\ifmmode#2\else$#2$\fi}}
\newcommand{\gVe}  {g^{\mathrm{e}}_{\mathrm{V}}}
\newcommand{\gAe}  {g^{\mathrm{e}}_{\mathrm{A}}}

\newcommand{\PZ}   {\mbox{$\mathrm{Z}$}}  
\definmath{\PWpm} {\mathrm{W}^{\pm}}      
\definmath{\Plp} {\ell^{+}}        
\definmath{\Plm} {\ell^{-}}        
\definmath{\Plpm}   {\ell^{\pm}}         
\definmath{\Pgtp} {\tau^{+}}        
\definmath{\Pgtm} {\tau^{-}}        
\definmath{\Pgtpm}   {\tau^{\pm}}         
\definmath{\Pgn}  {\nu}          
\definmath{\Pagn} {\overline{\nu}}     
\definmath{\Pf}      {\mathrm{f}}
\definmath{\Paf}  {\overline{\mathrm{f}}}
\definmath{\Pq}      {\mathrm{q}}
\definmath{\Paq}  {\overline{\mathrm{q}}}
\definmath{\Pu}      {\mathrm{u}}
\definmath{\Pau}  {\overline{\mathrm{u}}}
\definmath{\Pd}      {\mathrm{d}}
\definmath{\Pad}  {\overline{\mathrm{d}}}
\definmath{\Ps}      {\mathrm{s}}
\definmath{\Pas}  {\overline{\mathrm{s}}}
\definmath{\Pc}      {\mathrm{c}}
\definmath{\Pac}  {\overline{\mathrm{c}}}
\definmath{\Pb}      {\mathrm{b}}
\definmath{\Pab}  {\overline{\mathrm{b}}}
\definmath{\Pt}      {\mathrm{t}}
\definmath{\Pat}  {\overline{\mathrm{t}}}
\definmath{\Pap}  {\overline{\mathrm{p}}}
\definmath{\Pan}  {\overline{\mathrm{n}}}
\definmath{\PaD}  {\overline{\mathrm{D}}}
\definmath{\PaDz} {\overline{\mathrm{D}}^{0}}
\definmath{\PaB}  {\overline{\mathrm{B}}}
\definmath{\PaBz} {\overline{\mathrm{B}}^{0}}
\definmath{\PsDpm}   {\mathrm{D}^{\pm}_{\mathrm{s}}}  
\definmath{\PcgLpm}  {\Lambda^{\pm}_{\mathrm{c}}}  
\definmath{\PD} {\mathrm{D}}     
\definmath{\PDst} {\mathrm{D}^{*}}     
\definmath{\PgLz} {\Lambda^{0}}        

\newcommand{\massof}[1] {m_{\smash{#1}\mathstrut}}

\newcommand{\mPZ} {\massof{\mathrm{Z}}}

\newcommand{\mtautau}  {\massof{\mathrm{\tau\tau }}}
\newcommand{\GammaZ} {\Gamma_{\mathrm{Z}}}

\newcommand{\epem}   {\Pep\Pem}

\newcommand{\mumu}   {\Pgmp\Pgmm}
\newcommand{\tautau} {\Pgtp\Pgtm}

\newcommand{\ellell} {{\Plp}{\Plm}}

\newcommand{\nunu}   {\Pgn\Pagn}
\newcommand{\qqbar}  {\Pq\Paq}

\newcommand{\bbbar}  {\Pb\Pab}

\newcommand{\ZZ}{\mbox{\PZ\PZ}}
\newcommand{\zz}           {\PZ \PZ}
\newcommand{\eetozz}           {\epem \to \zz}
\newcommand{\eetozg}           {\epem \to \PZ \gamma^*}
\newcommand{\eetozzqqqq}       {\eetozz \to  \qqbar \qqbar}

\newcommand{\eetozzqqbb}       {\eetozz \to  \qqbar \bbbar}
\newcommand{\zztollll}         {{ \zz \to {\ellell} {\ellell}}}
\newcommand{\zztoqqll}         {{ \zz \to  \qqbar \ellell}}
\newcommand{\zztoqqee}         {{ \zz \to  \qqbar \epem}}
\newcommand{\zztoqqmm}         {{ \zz \to  \qqbar \mu^+ \mu^- }}

\newcommand{\zztobbll}         {{ \zz \to  \bbbar\ellell}}
\newcommand{\zztoqqnn}         {{ \zz \to  \qqbar \nunu}}
\newcommand{\zztobbnn}         {{ \zz \to  \bbbar \nunu}}
\newcommand{\zztollnn}         {{ \zz \to \ellell \nunu}}
\newcommand{\zztoqqqq}         {{ \zz \to \qqbar \qqbar}}
\newcommand{\zztoqqbb}         {{ \zz \to \qqbar \bbbar}}

\newcommand{\nnnn}       { \nunu \nunu}
\newcommand{\llnn}       { \ellell \nunu}
\newcommand{\eenn}       {\epem \nunu}
\newcommand{\mmnn}       {\mu^+ \mu^- \nunu}
\newcommand{\ttnn}       {\tau^+ \tau^- \nunu}
\newcommand{\llll}       {  \ellell \ellell}
\newcommand{\eeee}       { \epem \epem}
\newcommand{\eemm}       { \epem \mu^+ \mu^-}
\newcommand{\eett}       { \epem \tau^+ \tau^-}
\newcommand{\mmmm}       { \mu^+ \mu^-  \mu^+ \mu^-}
\newcommand{\mmtt}       { \mu^+ \mu^- \tau^+ \tau^-}
\newcommand{\tttt}       { \tau^+ \tau^- \tau^+ \tau^-}
\newcommand{\qqlnu}      { \mathrm{\qqbar} \ell \nu}
\newcommand{\qqll}       {  \qqbar \ellell}
\newcommand{\bbll}       { \bbbar \ellell}
\newcommand{\qqnn}       { \qqbar \nunu}
\newcommand{\bbnn}       { \bbbar \nunu}
\newcommand{\qqqq}       { \qqbar \qqbar}
\newcommand{\qqbb}       { \qqbar \bbbar}
\newcommand{\qqee}       { \qqbar \epem}
\newcommand{\bbee}        {\bbbar \epem}
\newcommand{\qqmm}       { \qqbar \mu^{+} \mu^{-} }
\newcommand{\bbmm}       { \bbbar \mu^{+} \mu^{-} }
\newcommand{\qqtt}       { \qqbar \tau^{+} \tau^{-}}
\newcommand{\bbtt}       { \bbbar  \tau^{+} \tau^{-}}
\newcommand{\llllz}      {\llll}
\newcommand{\qqeez}      {\qqee \, \& \, \overline{\bbee}}
\newcommand{\bbeez}      {\qqee \, \& \, \bbee      }
\newcommand{\qqmmz}      {\qqmm \, \& \, \overline{\bbmm}}
\newcommand{\bbmmz}      {\qqmm \, \& \, \bbmm      }
\newcommand{\qqttz}      {\qqtt \, \& \,  \overline{\bbtt}}
\newcommand{\bbttz}      {\qqtt \, \& \,  \bbtt      }
\newcommand{\xbttz}      {\bbtt \, \& \,  \overline{\qqtt}}
\newcommand{\qqnnz}      {\qqnn \, \& \,  \overline{\bbnn}}
\newcommand{\bbnnz}      {\qqnn \, \& \,  \bbnn      }

\newcommand{\qqqqz}      {\qqqq \, \& \,  \overline{\qqbb}}
\newcommand{\qqxbz}      {\qqbb \, \& \,  \overline{\qqqq}}
\newcommand{\qqbbz}      {\qqbb \, \& \,  \qqqq      }
\newcommand{\eennz}      {\eenn}
\newcommand{\mmnnz}      {\mmnn}
\newcommand{\ZZZ}[1]      { f_{#1}^{\mathrm{ZZZ}} }
\newcommand{\ZZG}[1]      { f_{#1}^{\mathrm{ZZ\gamma}} }

\newcommand{\roots} {\sqrt{s}}
\newcommand{\rootsp} {\sqrt{s'}}

\newcommand{\Ebeam}  {E_{\mathrm{b}}}
\newcommand{\Evis}   {\mbox{$E_{\mathrm{vis}}$}}

\definmath{\GeV}  {\mathrm{GeV}}
\definmath{\GeVc} {\mathrm{GeV}\!/c}
\definmath{\GeVcc}   {\mathrm{GeV}\!/c^2}
\definmath{\MeV}  {\mathrm{MeV}}
\definmath{\MeVc} {\mathrm{MeV}\!/c}
\definmath{\MeVcc}   {\mathrm{MeV}\!/c^2}
\definmath{\MVm}  {\mathrm{MV}\!/\mathrm{m}}
\definmath{\keV}  {\mathrm{keV}}
\definmath{\keVcm}   {\mathrm{keV}\!/\mathrm{cm}}
\definmath{\kV}      {\mathrm{kV}}
\definmath{\km}      {\mathrm{km}}
\definmath{\meter}   {\mathrm{m}}
\definmath{\cm}      {\mathrm{cm}}
\definmath{\mm}      {\mathrm{mm}}
\definmath{\micron}  {\mu\mathrm{m}}
\definmath{\nm}      {\mathrm{nm}}
\definmath{\kg}      {\mathrm{kg}}
\definmath{\gram} {\mathrm{g}}
\definmath{\second}  {\mathrm{s}}
\definmath{\microsec}   {\mu\mathrm{s}}
\definmath{\degree}  {^\circ}
\definmath{\degC} {^\circ\mathrm{C}}
\definmath{\ohm}  {\Omega}
\definmath{\Mohm} {\mathrm{M}\Omega}
\definmath{\rad}  {\mathrm{rad}}
\definmath{\mrad} {\mathrm{mrad}}
\definmath{\nb}      {\mathrm{nb}}
\definmath{\pb}      {\mathrm{pb}}
\newcommand{\eqref}[1]  {(\ref{#1})}
\newcommand{\PhysLett}  {Phys.~Lett.}

\newcommand{\PhysRev}   {Phys.~Rev.}
\newcommand{\NPhys}  {Nucl.~Phys.}
\newcommand{\NIM} {Nucl.~Instrum.\ Methods}

\newcommand{\CPC} {Comput. Phys. Commun.}
\newcommand{\EPJ} {Eur.~Phys.~J.} 
\newcommand{\OPALColl}    {OPAL Collab.}
\newcolumntype{L} {>{$}l<{$}}
\newcolumntype{C} {>{$}c<{$}}
\newcolumntype{R} {>{$}r<{$}}

\newcommand{\mee}{m_{\rm ee}}

\newcommand{\mmumu}{m_{\mu\mu}}

\newcommand{\mqq}    {m_{\mathrm{qq}}}

\newcommand{\pt}     {p_{\mathrm{t}}}
\newcommand{\pti}    {p_{\mathrm{t}i}}
\newcommand{\ptj}    {p_{\mathrm{t}j}}

\newcommand{\mvis}   {m_{\mathrm{vis}}}
\newcommand{\mll}    {m_{\ell \ell}}
\newcommand{\mrec}   {m_{\mathrm{recoil}}}

\newcommand{\LWW}    { {\cal L}_{\PW \PW}}
\newcommand{\Lqqnn}  { {\cal L}_{\qqnn}}

\newcommand{\Wenu}   { \PW  \Pe \nu}
\newcommand{\Leenn}  { {\cal L}_{\eenn}}
\newcommand{\Lmmnn}  { {\cal L}_{\mmnn}}
\newcommand{\xsec}  {\sigma_{\mathrm{ZZ}}}
\newcommand{\nexp}  {\mu_{\mathrm{e}}}
\newcommand{\nobs}  {n_{\mathrm{obs}}}
\newcommand{\nback} {n_{\mathrm{back}}}
\newcommand{\nsm}   {n_{\mathrm{ZZ}}}
\newcommand{\smtot} {n_{\mathrm{SM}}}
\newcommand{\eff}   {\epsilon_{\mathrm{chan}}}
\newcommand{\br}    {  B_{\mathrm{ZZ} }}
\newcommand{\lint}  { L_{\mathrm{int}} }
\begin{document}
%
%
\begin{titlepage}
%
\begin{center}
    \Large EUROPEAN ORGANIZATION FOR NUCLEAR RESEARCH
\end{center}
\bigskip 
\bigskip 
\begin{flushright}
    \large \PPEnum\\
    \Date \\
\end{flushright}
%
%
\begin{center}
    \huge\bf\boldmath

Study of Z Pair Production and  Anomalous Couplings
in $\mathrm{e^+ e^-}$ Collisions at $\sqrt{s}$ between
190\,GeV and 209\,GeV

\end{center}\bigskip\bigskip
\begin{center}{\LARGE The OPAL Collaboration
}\end{center}\bigskip\bigskip
\bigskip\begin{center}{\Large  Abstract}\end{center}
%
%

A study of Z-boson pair production 
in $\mathrm{ e^+ e^-}$ annihilation at center-of-mass
energies between
$190$\,GeV and  $209$\,GeV is reported.
Final states containing only leptons,
($\llll$ and $\llnn$),
quark and lepton pairs,
($\qqll$, $\qqnn$)
and only hadrons 
($\qqqq$) are considered.
In all states with at least one Z boson decaying
hadronically, lifetime, lepton and event-shape tags
are used to separate $\mathrm{ b \bar{b}}$ pairs from
$\mathrm{ q \bar{q}}$ final states.
Limits on  anomalous ZZ$\gamma$ and ZZZ couplings are
derived from the measured cross sections and from event 
kinematics using an optimal observable method.
Limits on low scale gravity with large extra dimensions are derived
from the cross sections and their dependence on polar angle.
\bigskip\bigskip\bigskip\bigskip
\bigskip\bigskip
\begin{center}
{\large Submitted to \EPJ  }
\end{center}
%
%
%
%
%
\end{titlepage}
\begin{center}{\Large        The OPAL Collaboration
}\end{center}
\begin{center}{
G.\thinspace Abbiendi$^{  2}$,
C.\thinspace Ainsley$^{  5}$,
P.F.\thinspace {\AA}kesson$^{  3}$,
G.\thinspace Alexander$^{ 22}$,
J.\thinspace Allison$^{ 16}$,
P.\thinspace Amaral$^{  9}$, 
G.\thinspace Anagnostou$^{  1}$,
K.J.\thinspace Anderson$^{  9}$,
S.\thinspace Arcelli$^{  2}$,
S.\thinspace Asai$^{ 23}$
% ...
}\end{center}
\vspace{-0.25cm}
$^{  1}$School of Physics and Astronomy, University of Birmingham,
Birmingham B15 2TT, UK
\newline
$^{  2}$Dipartimento di Fisica dell' Universit\`a di Bologna and INFN,
I-40126 Bologna, Italy
\newline
$^{  3}$Physikalisches Institut, Universit\"at Bonn,
D-53115 Bonn, Germany
\newline
% ...
$^{  5}$Cavendish Laboratory, Cambridge CB3 0HE, UK
\newline
% ...
$^{  9}$Enrico Fermi Institute and Department of Physics,
University of Chicago, Chicago IL 60637, USA
\newline
% ...
$^{ 16}$Department of Physics, Schuster Laboratory, The University,
Manchester M13 9PL, UK
\newline
% ...
$^{ 22}$Department of Physics and Astronomy, Tel Aviv University,
Tel Aviv 69978, Israel
\newline
$^{ 23}$International Centre for Elementary Particle Physics and
Department of Physics, University of Tokyo, Tokyo 113-0033, and
Kobe University, Kobe 657-8501, Japan
\newline
% ...
$^{  b}$ now at Physikalisches Institut, Universit\"at
Heidelberg, D-69120 Heidelberg, Germany
\newline
% ...
$^{  e}$ and Department of Experimental Physics, Lajos Kossuth University,
 Debrecen, Hungary
\newline
% ...
$^{  q}$ now at IPHE Universit\'e de Lausanne, CH-1015 Lausanne, Switzerland
\newline
% ...
$^{  s}$ now at Universitaire Instelling Antwerpen, Physics Department, 
B-2610 Antwerpen, Belgium
\newline
% ...

 
\section{Introduction}           \label{sec:intro}
%
%
LEP operation at center-of-mass 
energies above the $\PZ$-pair
production threshold has made a careful study of
the process $\eetozz$ possible. 
In the final data taking runs in 1999 and 2000 
LEP delivered an integrated luminosity of
more than 400~pb$^{-1}$ to the OPAL experiment at energies between
190\,GeV and 209\,GeV 
producing of the order of 100 detected $\PZ$-pair events.
In this paper we present results on $\PZ$-pair production from
these data.
Previously published studies from OPAL and the other 
LEP collaborations can be found in 
References~\cite{bib:zzopal,bib:zzaleph,bib:zzl3,bib:zzdelphi}.

\begin{figure}[b]
\center{
\epsfig{file=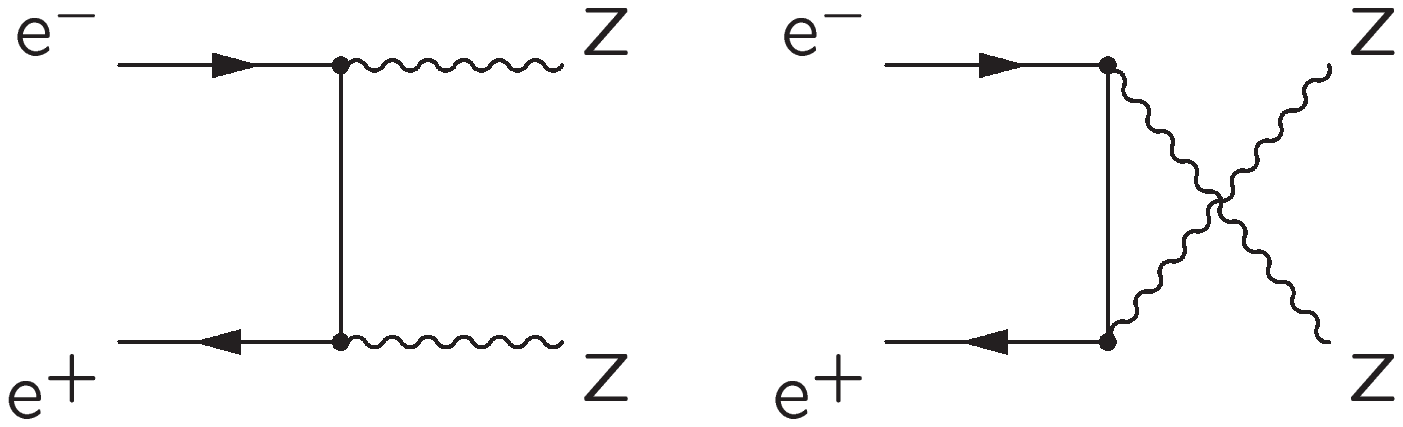,width=0.8\textwidth,
bbllx=40,bburx=595,bblly=400,bbury=650} }
\caption[$\eetozz$ ]{
\label{fig:eetozz}
NC2 Feynman diagrams for the process $\eetozz$ leading to
a final state with four fermions.
}
\end{figure}

In the Standard Model, the process $\eetozz$ occurs
via the NC2 diagrams~\cite{bib:zzfirst} shown in Figure~\ref{fig:eetozz}. 
The $\PZ$-pair cross section depends
on properties of the $\PZ$ boson (the $\PZ$ mass, $\mPZ$, 
the $\PZ$ resonance width, $\GammaZ$, and 
the vector and axial vector coupling of the $\PZ$ to 
electrons, $\gVe$ and $\gAe$) that have been measured with
great precision at the $\PZ$ resonance~\cite{bib:pdg}. 
The expected $\PZ$-pair cross section increases from about 
0.25~pb at \mbox{$\roots = 183$\,GeV} to slightly more than 1.0~pb above 
\mbox{$\roots = 200$\,GeV}, but remains more than an order of
magnitude smaller than the cross section for 
$\PW$-pair production.  
In contrast to $\PW$-pair production, where tree
level  $\PW\PW\gamma$ and $\PW\PW\PZ$ couplings are important, 
no tree-level  $\PZ\PZ\PZ$ and $\PZ\PZ\gamma$  couplings
are expected in the Standard Model. 
However, physics beyond the Standard Model could
lead to effective couplings~\cite{bib:hagiwara} 
which could then be observed
as deviations from the
Standard Model prediction
in the measured $\PZ$-pair cross section 
and kinematic distributions.
Such deviations have been proposed
in the context of two-Higgs-doublet models~\cite{bib:pal}
and in low scale gravity theories~\cite{bib:desh}.
In this paper we report on measurements of the NC2 $\PZ$-pair
cross section, including the extrapolation to final states 
where one or both $\PZ$ bosons have invariant masses far from
$\mPZ$.  
These measurements, together with optimal observables 
determined from the final states of
selected $\qqqq$ and $\qqll$ events,
are then used to extract limits
on possible $\PZ\PZ\PZ$ and $\PZ\PZ\gamma$ anomalous couplings.
Finally we use the measured cross sections, 
binned in polar angle,
to determine limits on large extra dimensions in low scale gravity
theories.

The analyses and presentation of this paper
largely follow that of the first OPAL $\eetozz$ results 
at center-of-mass energies of 
183\,GeV and 189\,GeV~\cite{bib:zzopal}.
All of the analyses have been optimized for the higher energies
and some adjustments to the analyses have been made to make them
less sensitive to background.

This paper is organized as follows.
In Section~\ref{sec:data} we describe the data
used and the Monte Carlo simulation
of signal and background processes.
In Section~\ref{sec:sel} we 
describe the selections for the processes
$\zztollll$, $\zztollnn$, $\zztoqqnn$,
$\zztoqqll$ and $\zztoqqqq$,
where $\ellell$ denotes a charged lepton pair of opposite 
charge and $\qqbar$ any of the five lightest quark-antiquark pairs.
We also consider  the processes $\zztobbll$,
$\zztobbnn$ and $\zztoqqbb$ 
using b-tagging methods similar to those developed for
the OPAL Higgs search~\cite{bib:opal-higgs}.
In Section~\ref{sec:xsec} the selected events 
are used to measure the $\PZ$-pair cross section.
In this section we also test our measured values of the
$\PZ$-pair cross section and the value of the 
branching ratio BR$(\PZ \to \bbbar)$ 
measured in $\PZ$-pair decays
for consistency with the Standard Model prediction.
At LEP,
$\PZ$-pair events with b quarks in the final state form 
the principle
background to the search for Higgs bosons produced in association with a Z
boson. 
In Section~\ref{sec:ac} 
anomalous neutral-current triple gauge couplings
are constrained 
using the measured cross sections and kinematic
variables combined in an optimal observables (OO) method.
Finally, in Section~\ref{sec:ed} limits are placed on low scale
gravity theories with large extra dimensions.

\section{Data samples and Monte Carlo simulation}        \label{sec:data}

The OPAL detector\footnote{OPAL uses a right-handed coordinate system in
which the $z$ axis is along the electron beam direction and the $x$
axis points towards the center of the LEP ring. 
The polar angle, $\theta$, is measured with respect
to the $z$ axis and the azimuthal angle, $\phi$, with respect to the
$x$ axis.}, trigger and data acquisition system 
used for this study are described in 
References~\cite{bib:OPAL-detector,bib:OPAL-SI,bib:OPAL-SW,bib:OPAL-TR,
bib:OPAL-DAQ}.
 
The approximate integrated luminosities and luminosity-weighted mean
center-of-mass energies~\cite{bib:ELEP} for
the data used in this analysis 
are given in Table~\ref{tab:lumi}.
The actual integrated
luminosities used for each final state vary due to differing 
detector status requirements.
These luminosities were measured using small-angle Bhabha scattering events
recorded in the silicon-tungsten 
luminometer~\cite{bib:OPAL-SW,bib:OPAL-SM172,bib:OPAL-SM183}
and the theoretical calculation given in Reference~\cite{bib:bhlumi}.
The overall uncertainty on the luminosity measurement amounts to 
less than 0.5\% and contributes negligibly to our 
cross-section measurement error. 

\begin{table}
\begin{center}
\begin{tabular}{|c|c|c|}
\hline
Energy point label &
Mean center-of-mass energy &  
Approximate integrated luminosity    \\
(GeV)  &
(GeV)  &
(pb$^{-1}$) \\
\hline
192 & \emnc & \lumic \\
196 & \emnd & \lumid \\
200 & \emne & \lumie \\
202 & \emnf & \lumif \\
205 & \emng & \lumig \\
207 & \emnh & \lumih \\
\hline
\end{tabular}
\end{center}
\caption{ Luminosity-weighted mean
center-of-mass energies and typical 
integrated luminosities for
the data presented here. The 
integrated luminosities for  each channel
vary slightly since they depend on the status of
different detector elements.  The channel dependent luminosities
are given in  Tables~\ref{tab:zzxsec192} - \ref{tab:zzxsec207}.
\label{tab:lumi}
}
\end{table}

Selection efficiencies and backgrounds were calculated 
using a
simulation~\cite{bib:gopal} of the OPAL detector.
The simulated events were 
processed in the same manner as the data.
We define the $\PZ\PZ$ cross section as the contribution to
the total four-fermion cross section from the NC2
$\PZ$-pair diagrams
shown in Figure~\ref{fig:eetozz}.
All signal efficiencies given in this paper are with
respect to these $\PZ$-pair processes.
Contributions from all other four-fermion final states,
including interference with NC2 diagrams,
are considered as background.
For studies of the signal efficiency
we have used grc4f~\cite{bib:grc4f} and
YFSZZ~\cite{bib:yfszz}  with
PYTHIA~\cite{bib:pythia} used for the parton shower
and hadronization.
YFSZZ only generates the NC2 diagrams.  
The grc4f Monte Carlo includes four-fermion
background processes and the interference
of background and the NC2 $\PZ\PZ$ signal.  
Weights, based on the grc4f matrix 
elements, are assigned to each grc4f Monte Carlo
event for the event to originate from
NC2 $\PZ\PZ$ signal, four-fermion background and
interference between the four-fermion background
and the NC2 $\PZ\PZ$ signal.  In all of the fits 
performed in this analysis we use the
ZZTO calculation~\cite{bib:mc4fermi} with 
the $G_F$ renormalization scheme for 
integrated Standard Model cross sections.
To check the simulation of anomalous couplings in
the YFSZZ Monte Carlo we use EEZZ~\cite{bib:baur}.

Backgrounds are 
simulated using several different generators.
KK2f~\cite{bib:kk2f}
(with PYTHIA used for the parton shower and hadronization)
is used to simulate
two-fermion final states such as 
$\epem \to \PZ^* (n\gamma) \to \qqbar (n\gamma) $ and 
$\epem \to \gamma^* (n\gamma) \to \qqbar (n\gamma) $,
where $(n\gamma)$ indicates the generation of one or
more initial-state radiated
photons. HERWIG~\cite{bib:herwig} and PYTHIA
are used as checks for these final states.
These two-fermion samples include gluon radiation
from the quarks, which produce $\qqbar {\rm g}$, 
$\qqbar \qqbar$ and
$\qqbar{\rm gg}$ final states.
KK2f is also used to simulate muon and tau pair events
and Bhabhas are modeled using BHWIDE~\cite{bib:bhwide} and
TEEGG~\cite{bib:teegg}.
The grc4f generator, with the contribution 
exclusively due to NC2 diagrams removed,
is used to simulate  other four-fermion background.
KORALW~\cite{bib:koralw} and EXCALIBUR~\cite{bib:excalibur} are
used as checks of the four-fermion background.

Multiperipheral (``two-photon'') processes with hadronic final
states are simulated by
combining events from PHOJET\cite{bib:phojet}, 
for events without electrons\footnote{In this paper reference to
a specific fermion, such as an electron, also includes
the charge conjugate particle, in this case the positron.}
scattered
into the detector, and
HERWIG~\cite{bib:herwig},
for events with single electrons scattered
into the detector.
Two-photon events with both the electron and positron
scattered into the detector, which are only
a signficant background for
the $\qqee$ final state, are simulated with
TWOGEN~\cite{bib:twogen}.
The Vermaseren~\cite{bib:vermaseren} generator is
used to simulate
multiperipheral production of the final states $\epem \ellell$.

To avoid background from 
four-fermion final states mediated by
a $\PZ$ boson and a virtual $\gamma^*$,
our selections were optimized on Monte Carlo samples
to select events with candidate $\PZ$-boson masses,
$m_{1}$ and $m_{2}$, that satisfy
\mbox{$ m_{1} + m_{2}     > 170 \ \mathrm{GeV}$} and 
\mbox{$| m_{1} - m_{2} |  <  20 \ \mathrm{GeV}$}.
Above 190\,GeV
more than 90\% of the events
produced via the NC2 diagrams are contained in this mass
region.
Events from the 
NC2 diagrams dominate in this mass region except for
final states containing electron pairs.
However, inside the acceptance of the electromagnetic 
calorimeters\footnote{
The acceptance of the electromagnetic calorimeters for
electrons is approximately given by $| \cos \theta_{\mathrm{e}} |  <  0.985$,
where  $\theta_{\mathrm{e}}$ is the polar angle
of the electron.},
the backgrounds from two-photon
and electroweak Compton scattering
($\Pe \gamma \to \Pe \PZ$) 
processes~\cite{bib:zee} 
to $\PZ$ final states with one $\PZ$ decaying to
electron pairs
is reduced to less than 5\% of the expected $\PZ$-pair cross section
and the NC2 diagrams again dominate.
\section{Event selection}           \label{sec:sel}

The OPAL
selections cover all $\PZ\PZ$ final states except
$\nnnn$ and $\ttnn$. 
In hadronic final states, the energies and directions of the jets
are determined using tracks to reconstruct charged particles
and electromagnetic and hadronic calorimeter clusters
to reconstruct neutral particles.
The correction for unavoidable double counting due to the
overlap of calorimeter energy deposited from charged and neutral particles is
described in Reference~\cite{bib:opal-higgs}.

In the $\qqee$, $\qqmm$ and $\qqqq$ analyses,  
four-constraint (4C) and five-constraint (5C) kinematic fits are used.
The 4C fit imposes energy and momentum conservation.
In the 5C fit the added constraint requires 
the masses  of the two candidate $\PZ$ bosons to be equal
to one another.
For final states with either $\PZ$ boson decaying
to a tau pair, the direction of each tau lepton
is approximated by the reconstructed particles
that identify the tau. 
The energies and total momenta of the
tau leptons are obtained by leaving the reconstructed
direction of the four fermions fixed and scaling
the energy and momentum of each of the fermions
to obtain energy and momentum conservation.  The
scaled values of the tau momentum and energy
are then used in the subsequent steps of the analysis.
In the
$\qqtt$ final state, subsequent kinematic
fits are therefore effectively 2C and 3C fits.

The selection procedures for events containing charged leptons or neutrinos
($\llll$,$\llnn$, $\qqll$ and $\qqnn$) are largely unchanged
from the analysis used at 183\,GeV and 189\,GeV in Reference~\cite{bib:zzopal}, 
except that the likelihood functions and some of the
cuts have been optimized
for the higher energies.  More significant changes have been made 
to the $\qqqq$ and $\qqbb$ selections.

\subsection{Selection of {\mybold $\zztollll$} events }
\label{ssec:llll}

$\PZ$-pairs decaying to final states 
with four charged leptons ($\llll$)
produce low multiplicity events with
a clear topological signature that
is exploited to
maximize the selection efficiency.
The $\llll$ analysis begins by
selecting low multiplicity events (less than
13 reconstructed good tracks and less than 
13 electromagnetic clusters) with visible
energy of at least $0.2 \roots$ and at least
one good track with momentum
of 5\,GeV or more.  Good tracks are required to be 
consistent with originating from the interaction point
and
to be composed of space points from 
at least half of the maximum possible number of 
central tracking detector (Jet Chamber) sense wires.

Using a cone algorithm,
the events are required to  have 
exactly four cones of $15^\circ$ half angle
each containing between 1 and 3 tracks.
Cones of opposite charge are paired\footnote{Two-track 
cones are assigned the charge of 
the more energetic track if
the momentum of one track exceeds that of
the other by a factor of 4.  
Events with
a cone which fails this requirement are rejected.}
to form $\PZ$ boson candidates.

Lepton identification is
only used to classify events as background 
or to reduce the number of cone combinations
considered by preventing the matching of
identified electrons with identified muons.
Electrons are identified on the basis of energy
deposition in the electromagnetic calorimeter,
track curvature and
specific ionization in the tracking chambers.  Muons
are identified using the association
between tracks and 
hits in the hadron calorimeter and muon chambers.

To reduce background from two-photon events with a single
scattered electron detected, we eliminate events
with forward-going electrons or backward-going positrons
with the cuts \mbox{$ \cos \theta_{\mathrm{e^-}} < 0.85$} 
and \mbox{$ \cos \theta_{\mathrm{e^+}} > -0.85$}.
Here $\theta_{\mathrm{e^-}}$ (  $\theta_{\mathrm{e^+}}$)
is the angle of the electron (positron) with
respect to the incoming electron beam.
Background from
partially reconstructed $\qqbar (n\gamma)$  
events and two-photon events is reduced by
requiring that most of the energy 
not be concentrated in a single cone.
Defining $E_{\mathrm{vis}}$ as the total visible energy of the
event and $E_{\mathrm{cone}}^\mathrm{max}$ as the energy
contained in the most energetic cone we require
$E_{\mathrm{vis}} -  E_{\mathrm{cone}}^\mathrm{max} > 0.2\roots$.

The invariant masses of the lepton pairs are calculated in
three different ways which are motivated by the 
possibility of having zero, one  or two tau pairs 
in the event.
The events are first classified according to
the assumed number of tau pairs in the event:
\begin{enumerate}
\item Events without tau pairs can be selected by requiring
a high visible energy.  Therefore,
events with  $\Evis > 0.9 \roots$ 
are treated as
$\eeee$, $\eemm$ or $\mmmm$ events. 
We also treat all events 
with $| \cos \theta_{\mathrm{miss}}|  > 0.98$
($\theta_{\mathrm{miss}}$ is the polar angle
associated with the missing momentum in the
event)
as $\eeee$, $\eemm$ or $\mmmm$ events to
maintain efficiency for $\PZ$-pairs with
initial-state radiation.
As there are no missing neutrinos in these events,
the mass  of each cone-pair combination is evaluated
using the measured energies and momenta of the leptons.
 
\item  Events failing (1) with a cone-pair combination
that has energy exceeding  $0.9 \mPZ$ are tried
as an $\eett$ or $\mmtt$ final state. 
The mass of the tau-pair system
is calculated from the recoil mass of the presumed
electron or muon pair.

\item Events failing (1) with a cone-pair combination
failing (2) are treated as
$\tttt$ final states. The momenta of the
tau leptons are determined using the scaling procedure
described in the introduction to Section~\ref{sec:sel}.
The invariant masses of the cone
pairs
are evaluated using the scaled momenta.  
\end{enumerate}

\begin{figure}
  \center{
   \begin{tabular}{cc}    
     \epsfig{file=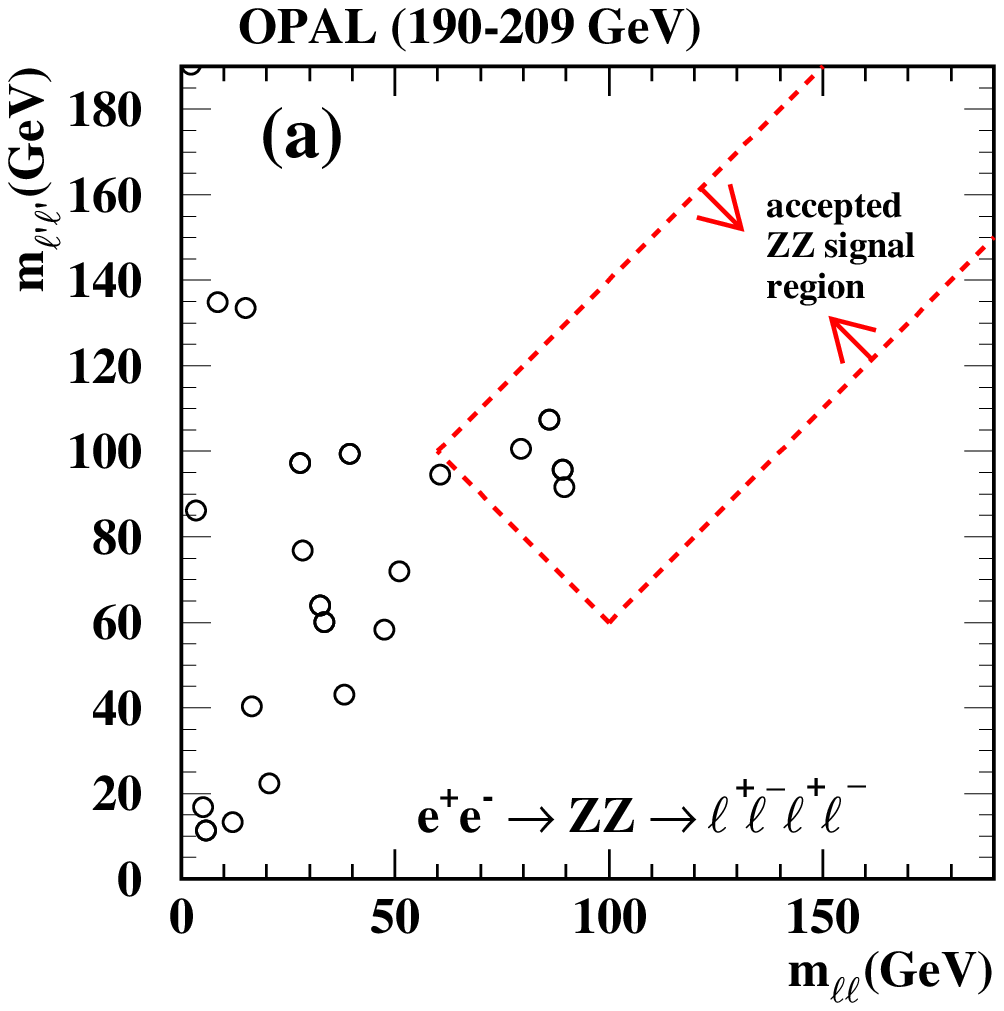,width=0.45\textwidth`
      } &
    \epsfig{file=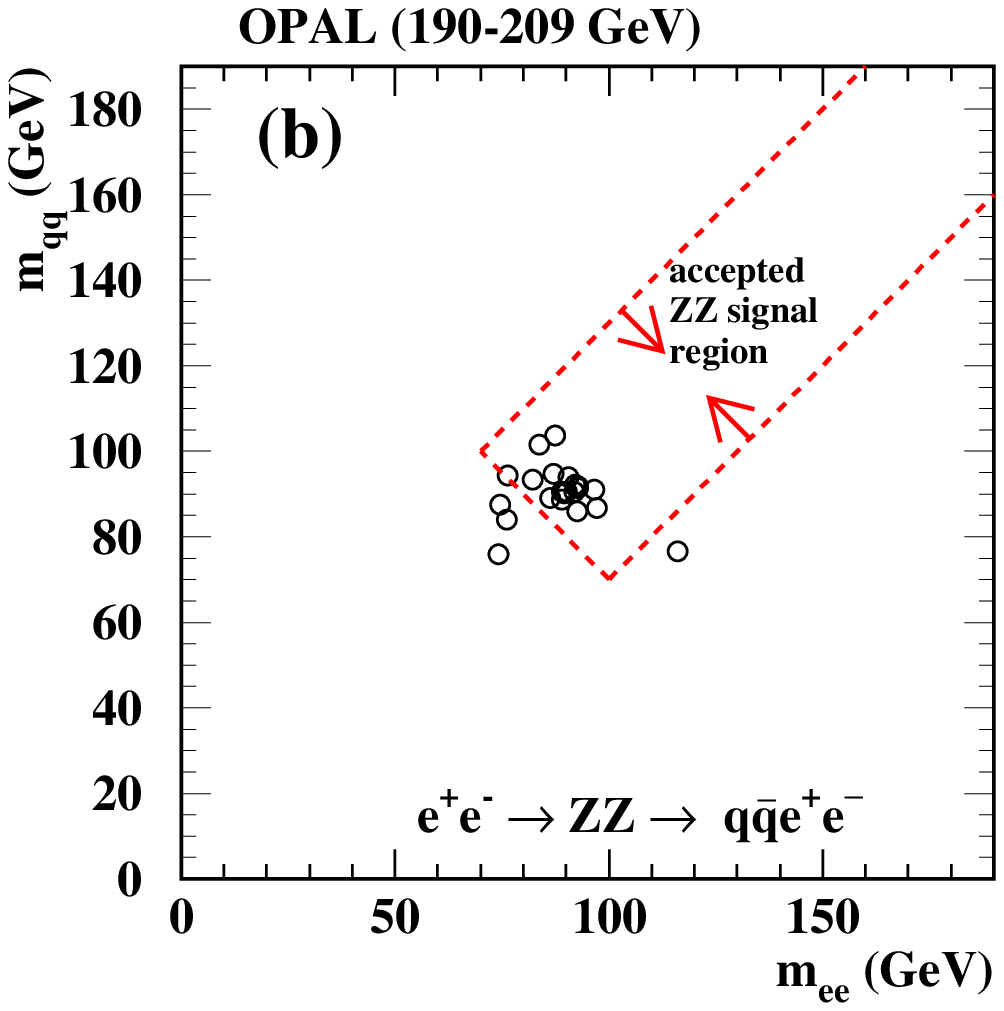,width=0.45\textwidth
      }\\    
    \epsfig{file=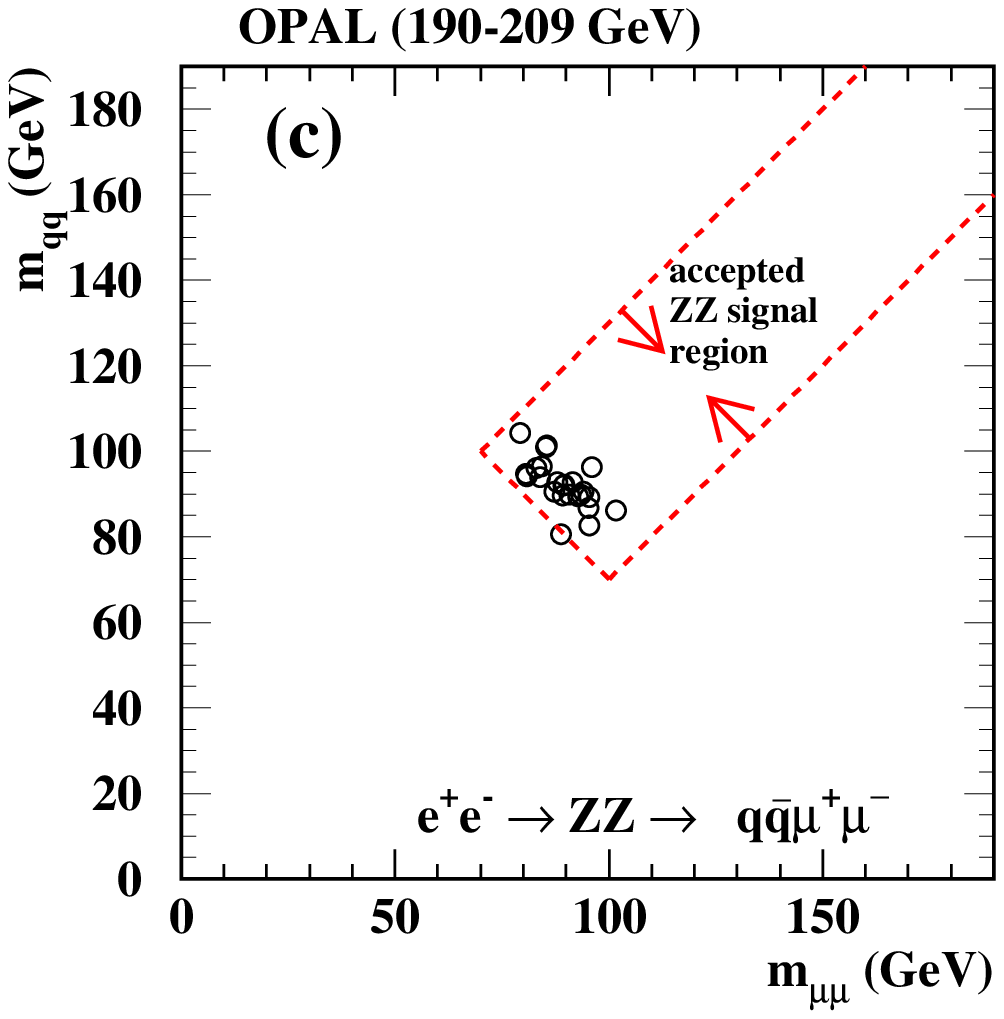,width=0.45\textwidth
      } &
    \epsfig{file=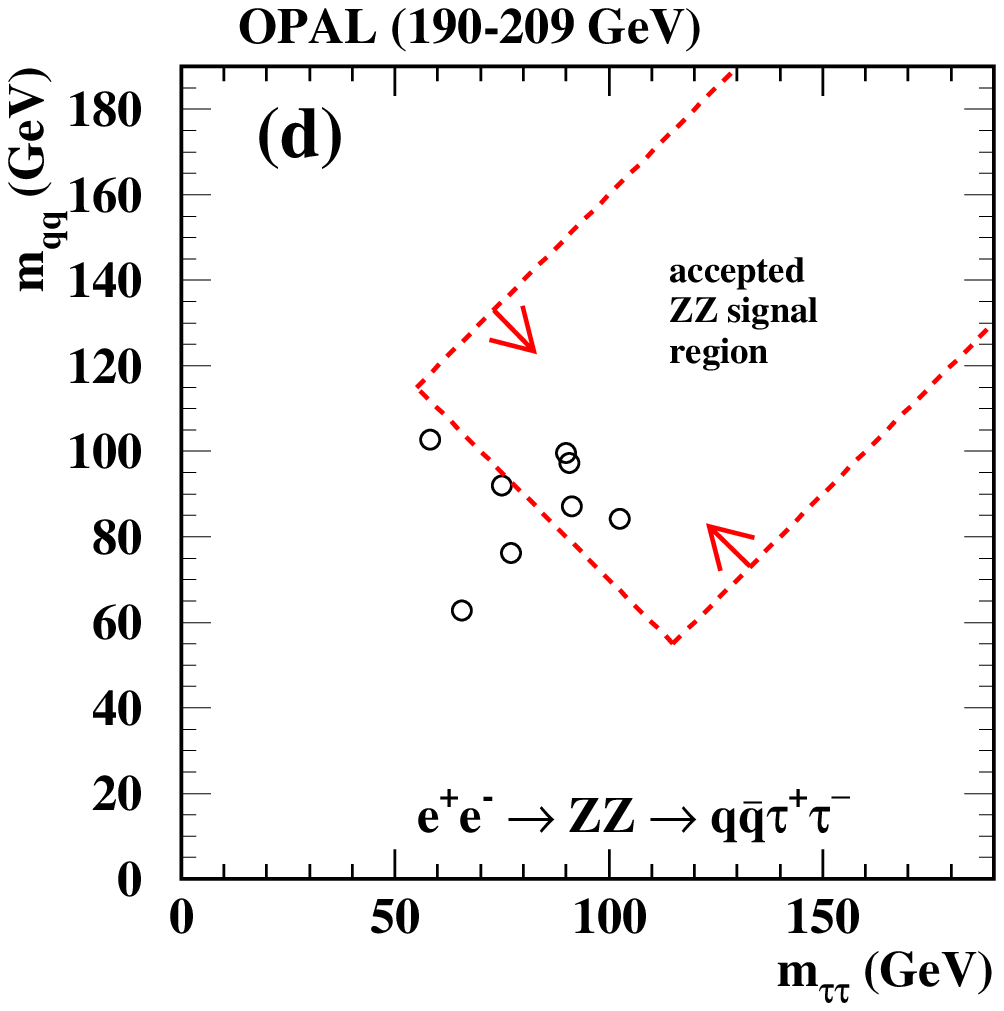,width=0.45\textwidth
      }\\
   \end{tabular}
}
  \caption[ zzllll ]{
  \label{fig:zzmm}
(a) Invariant masses of $\llll$ cone pairs.
Invariant mass pairs for the 
(b) $\qqee$ 
(c) $\qqmm$  and 
(d) $\qqtt$ data.
The dashed lines show the final invariant mass cuts.
In contrast to Reference~\cite{bib:zzopal}, 
the $\qqee$, $\qqmm$ unconstrained invariant masses 
are plotted after the requirement
that the 5C fit mass, with both quark and lepton-pair
masses constrained to be the same, exceeds 85\,GeV.  
The $\qqtt$  unconstrained invariant masses 
are shown after a similar cut on the 3C fit mass.
}
\end{figure}

In any event where the lepton identification
allows more than one valid combination, 
each combination is tested
using invariant mass cuts.
For events with one or more combinations of cone-pairings
satisfying 
$|\mPZ - m_{\ell\ell}| < 0.1 \mPZ$ 
and $|\mPZ - m_{\ell'\ell'}| < 0.1 \mPZ$
the one with the smallest
value of $(\mPZ - m_{\ell\ell} )^2 + (\mPZ - m_{\ell'\ell'} )^2$
is selected for further analysis.
In the other events, the combination with 
the smallest value of $| \mPZ -  m_{\ell\ell} |$ or
$| \mPZ -  m_{\ell'\ell'} |$ is selected.
These requirements maintain efficiency for signal
events with a single off-shell $\PZ$ boson.
The final event sample is then chosen with
the requirement
$m_{\ell\ell} + m_{\ell'\ell'}  >  160$\,GeV and
$| m_{\ell\ell} - m_{\ell'\ell'} |   <  40$\,GeV.
The signal
detection efficiency, averaged over all $\llll$ final
states is approximately 56\%.
The efficiency for individual final states ranges
from about 30\% for $\tttt$ to more than 70\% for $\mmmm$.
These efficiencies have almost no dependence on center-of-mass
energy.
In Table~\ref{tab:zzxsum} (line~\lllllb)
we give the efficiency, background
and observed number of events.
The errors on the efficiency and background include the systematic
uncertainties which are discussed in Section~\ref{ssec:sys}. 
The invariant masses of all cone pairs passing one of the
selections are shown in 
Figure~\ref{fig:zzmm}a.
A total of four $\llll$ events is  observed between 190\,GeV and 209\,GeV
with an expected background of $1.08\pm0.27$ events.  The 
remaining background
is dominated by four-lepton events which are not from the
NC2 diagrams that define $\PZ$-pair production.

%
%

\subsection{Selection of {\mybold $\zztollnn$}\ events }
\label{ssec:llnn}
The selection of the $\eenn$ and $\mmnn$ final states
is based on the OPAL selection of
$\PW$ pairs decaying to
leptons~\cite{bib:OPAL-WW}.
The mass and momentum
of the  $\PZ$ boson decaying to $\nunu$
are calculated using the 
beam-energy constraint and the 
reconstructed energy and momentum 
of the $\PZ$ boson decaying to a charged lepton pair.
A likelihood selection based on the visible and recoil masses
as well as the polar angle of the leptons,
is then used to separate signal from background.
 
The $\eenn$ selection starts with OPAL
$\PW$-pair candidates where both charged
leptons are classified as electrons.
Each event is
then divided into two hemispheres using the thrust axis.
In each hemisphere, the track with the highest momentum
is selected as the leading track.
The sum of the charges of these two tracks is required to be zero.
The determination of the 
visible mass, $\mvis$, and the recoil mass,
$\mrec$, is based on the
energy as measured in the electromagnetic calorimeter and the direction
of the leading tracks.

\begin{figure}
\begin{center}
\epsfig{file=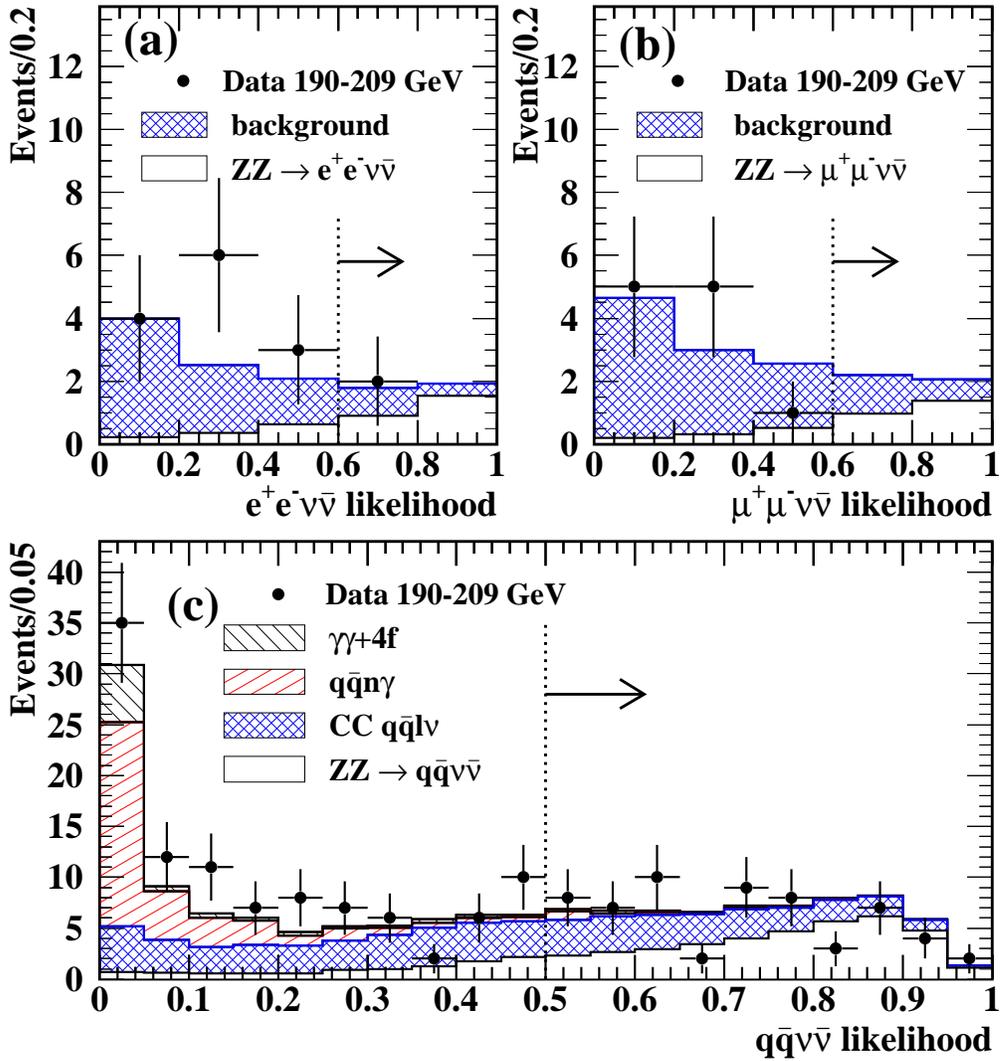,width=0.95\textwidth}
\end{center}
\vspace{-1.0cm}
\caption[$\qqnn$ likelihood ]{
Likelihood discriminant
used at $\roots > 190$\,GeV for 
(a) the $\eenn$ selection,
(b) the $\mmnn$ selection and the
(c) the $\qqnn$ selection.
The dotted line and arrow show the position of the likelihood cut used
to select ZZ events in each case.
\label{fig:zzlikenn}
}
\end{figure}

The likelihood selection uses three variables: 
$Q\cos\theta$, 
where $\theta$ is the angle of the track 
with the highest momentum  and 
$Q$ is its charge,
the normalized sum of visible and recoil masses $(\mvis+\mrec)/\roots$
and 
the difference of visible and recoil masses, $\mvis-\mrec$.
The performance of the
likelihood selection is improved with the following preselection:
$\mathrm{-25 \,GeV} < \mvis-\mrec < \mathrm{15 \,GeV}$ and
$       (\mvis+\mrec)> 170 $\,GeV.
Two events with $\Leenn >0.60$ are selected.
See Table~\ref{tab:zzxsum} (line~\eennlb) and Figure~\ref{fig:zzlikenn}a.
The expected background is primarily from $\PW$ pairs
and amounts to $1.28 \pm 0.15$ events.

The $\mmnn$ selection starts 
with the OPAL $\PW$-pair candidates
where both charged leptons are
classified as muons.
The selection procedure is  the same as for the $\eenn$
final states except that $\mvis$ and $\mrec$
are calculated from the  momentum of the reconstructed tracks
of the $\PZ$ boson decaying to muon pairs.  
The
likelihood preselections
$\mathrm{-25 \,GeV} < \mvis-\mrec < \mathrm{25 \,GeV}$ and
$(\mvis+\mrec)> 170 $\,GeV are applied.
No event with $\Lmmnn >0.60$ is selected while $4.30 \pm 0.39$ are
expected
(see Table~\ref{tab:zzxsum} (line~\mmnnlb) and Figure~\ref{fig:zzlikenn}b).
The probability to observe no event when $4.3 \pm 0.39$ are expected
is approximately 1.4\%.

\subsection{Selection of {\mybold $\zztoqqll$}\ events}
\label{ssec:qqll}
The  lepton pairs in the $\qqbar\epem$\ 
and $\qqbar\mumu$\ final states
have a distinctive signature 
making possible selections with high 
efficiencies and low background contaminations.  
In the $\qqtt$ final state,
the decay of the tau leptons produces
events which are more difficult to identify.
The identification of this final state exploits
the missing momentum and
missing energy carried away by the neutrinos
produced in the decay of the tau lepton.   In comparison 
to our previous publication~\cite{bib:zzopal}, the 
$\qqbar\epem$\ and $\qqbar\mumu$\ selections are almost unchanged, but
the analysis of $\qqtt$ events with b-tags has been simplified.
The $\qqee$ and $\qqmm$ selections presented here are also similar to
those used in \cite{bib:opal-4f} but are optimized specifically for
$\PZ$-pair final states.

\subsubsection{Selection of {\mybold $\zztoqqee$}\ 
and  {\mybold $\zztoqqmm$} events}

The selection of $\qqee$ and $\qqmm$ final states 
requires the visible
energy of the events
to be greater than 90\,\mbox{GeV} and at least six 
reconstructed tracks. 
Among all tracks with
momenta greater than 2\,GeV, the highest momentum track is taken as
the first lepton candidate and the second-highest momentum track
with a charge opposite to the first candidate is taken as the second
lepton candidate. 
Using the Durham~\cite{bib:durham} jet
algorithm, the event, including the lepton candidates, 
is forced into four jets  
and the jet  resolution variable that separates the three-jet
topology from the four-jet topology,
$y_{34}$, is required to be greater than $10^{-3}$. 
Excluding the electron or muon candidates
and their associated calorimeter clusters, the rest of the event is
forced into two jets.  The 4C and 5C fits to
the two lepton candidates and the two jets are
required to converge.\footnote{ In the context of
this paper, convergence is defined as a
fit probability greater than $10^{-10}$ after at
most 20 iterations.}

In the $\qqee$ selection 
no explicit electron identification
is used. Electron candidates are selected by 
requiring the sum of the electromagnetic cluster
energies $E_1 + E_2$ associated to the electrons to be greater than
70\,GeV and the momentum of the track 
associated with the most energetic electron candidate
to exceed 20\,GeV.
We also reject the event if
the angle between either electron candidate 
and any other track is less than 5$^\circ$.  

In the $\qqmm$ selection the muons are identified using
(i) tracks which match a reconstructed segment in
the muon chambers,
(ii) tracks which are associated to hits in the hadron calorimeter
or muon chambers~\cite{bib:OPAL-WW},
or 
(iii) isolated tracks associated to
electromagnetic clusters with reconstructed energy less than 2\,GeV.
No isolation requirement is
imposed on events with both muon tracks passing (i) or (ii).
Events with at least one muon identified with criterion (iii)
are accepted if both muon candidates in the
event have an angle of at least 10$^\circ$ to the 
nearest track.
We require the sum of the momenta of the
two leptons to be greater than 70\,GeV.

$\PZ$-pair events are separated from $\PZ\gamma^*$ background
by requiring
the fitted mass of the 5C fit
to be larger than 85\,GeV
and the invariant masses $\mll$ and $\mqq$ obtained
from the 4C fit to satisfy
 $ (\mll + \mqq)    >  170$\,GeV and
 $|\mll - \mqq |  <   30$\,GeV.
Figure~\ref{fig:zzmm}b (\ref{fig:zzmm}c) shows the
distribution of $\mee$\ ($\mmumu$) and $\mqq$\ before the cuts on the
masses from the 4C fit.

After all cuts the mean selection efficiency\footnote{
Small amounts of feedthrough from other $\PZ\PZ$ final states, in
this case $\qqtt$, are counted as signal.} 
for
$\qqee$
signal events is 
60\% and depends weakly on energy.
A total of 17 candidate events is found after all cuts at 
$\sqrt{s}$ between 190\,GeV and 209\,GeV, which can be 
compared to the Standard Model expectation 
(including a small background) of $14.1 \pm  0.4$.
In Table~\ref{tab:zzxsum} (lines~\qqeelb\ and \bbeelb)
we give the efficiency, background
and observed number of events.
The errors on these efficiencies and backgrounds include the systematic
uncertainties (see Section~\ref{ssec:sys}). 
The largest source of background after all
cuts is from  the process  $\eetozg\to\qqee$\ and from two-photon
events.

In the $\qqmm$ selection, the mean selection efficiency is 
73\%
and varies less than 5\% with energy.
A total of
22 events is observed at
$\sqrt{s}$ between 190\,GeV and 209\,GeV, which can be 
compared to the Standard Model expectation 
(including a small background) of $15.5 \pm  0.4$.
(See Table~\ref{tab:zzxsum} lines~\qqmmlb\ and \bbmmlb).
The  background after all cuts is expected to come
mainly from $\eetozg\to\qqmm$\ events.

\subsubsection{Selection of {\mybold $\eetozz\to\qqbar\tautau$}\ events}

The $\qqtt$ final state is selected from a
sample of events with track
multiplicity greater than or equal to six.  
Events which 
pass all the requirements of the 
$\qqee$ and $\qqmm$ selections
are
excluded from this selection.
The tau-lepton
candidates are selected using a neural network algorithm
which is described in detail in Reference~\cite{bib:opal-higgs}.  
At least two tau candidates are required.
The tau candidate with 
the highest neural network output value is taken as the
first candidate. The second candidate is required to have its
charge opposite to the first candidate and the highest 
neural network output
value among all remaining candidates.  

Because of the presence of neutrinos in the final state
the missing energy, $\roots - E_{\mathrm{vis}}$, is
required to exceed 15\,GeV  while
the visible energy of the event, $E_{\mathrm{vis}}$,  
is required to exceed 90\,GeV.
In addition, the sum of the momenta of the leading
tracks from the tau-lepton 
decays is required to be less than 70\,GeV. 
Since the direction of the
missing momentum in signal events will tend to be along the direction
of one of the decaying tau leptons, the angle
$\alpha_{\tau,{\rm miss}}$ between the missing momentum and a 
tau-lepton
candidate is required to satisfy $\alpha_{\tau,{\rm miss}} < 90^\circ$ 
for at
least one of the two tau candidates. 

The two hadronic jets are selected in
the same way as in the $\eetozz\to\qqbar\epem$ selection.
The 
initial estimate of the energy and the momenta of the tau
candidates is found from the sum of the
momenta of the tracks associated to the tau by the
neural network algorithm and all unassociated electromagnetic clusters
in a cone with a half angle of 10$^\circ$ around the leading 
track from the tau decay.  
A 2C kinematic fit that imposes energy and momentum
conservation (see the introduction to
Section~\ref{sec:sel}) is required to converge.
A 3C kinematic fit, with the additional constraint 
of the equality of the fermion pair masses is also  required to converge.

Using the network output~\cite{bib:opal-higgs} for each
tau candidate, a tau lepton 
probability, ${\cal P}$, is calculated taking into account the different
branching ratios, sensitivities, efficiencies and background levels
for 1-prong and 3-prong tau-lepton decays.
In this paper, we
combine two probabilities 
${\cal P}_1$ and ${\cal P}_2$,
to form a likelihood ratio using
\begin{equation}
{\cal L} = \frac{{\cal P}_{1}{\cal P}_{2}}
{{\cal P}_{1}{\cal P}_{2} + (1-{\cal P}_{1})(1-{\cal P}_{2})}  .
\label{eqn:ttprob}
\end{equation}
The likelihood ratio associated with probabilities of the
two tau candidates is required to 
satisfy ${\cal L}_{\tau\tau} > 0.977$.
In addition, the common mass of the 3C fit is required to
exceed 85\,GeV.
Using the 2C fit masses of the tau pair, $\mtautau$, and the 
quark pair, $\mqq$, as obtained from the kinematic fit, we also require 
$\mqq + \mtautau   > 170$\,GeV and
$| \mqq - \mtautau | < 60$\,GeV.

Four candidate events are found in the 
data.
Figure~\ref{fig:zzmm}d shows the masses of the candidate
events before the invariant mass cuts.
In Table~\ref{tab:zzxsum} (lines~\qqttlb\ and \bbttlb)
we give the efficiencies, backgrounds
and observed number of events.

\subsubsection{Selection of {\mybold $\eetozz\to \bbll$} }

The selection of the $\bbll$ events is based on the $\qqll$
selections with addition of the algorithm
described in Reference~\cite{bib:opal-higgs} to identify
$\bbbar$ final states.
The probabilities that each of the hadronic jets
is a b jet
can be combined to form a
likelihood function, ${\cal L}_{\mathrm{bb}}$,  
according to Equation~(\ref{eqn:ttprob}).
Because the 
$\qqee$ and 
$\qqmm$ selections are pure, a relatively loose cut
of $ {\cal L}_{\mathrm{bb}} > 0.2$ is used to
select the $\bbee$ and $\bbmm$ samples.
For the selections with electron and muon pairs 
there are two classes of events since the selected
$\bbll$ events are a subset of the 
$\qqll$ events.
In Table~\ref{tab:zzxsum} (lines~\bbeelb\ and \bbmmlb)
we give the efficiencies of the
b-tagged samples with respect to the expected fraction of 
$\bbbar$ events.  The efficiencies for samples without b-tags
are given with respect to the hadronic decays
without $\bbbar$ final states.

In the data with $\sqrt{s}$ above 190\,GeV
we find three candidate  $\bbee$\ events
with $2.49 \pm 0.19$ expected and  
in the $\bbmm$\ selection we find seven events with $2.59 \pm 0.15$
events expected.
The probability to observe seven or more
events when $2.59\pm 0.15$ are expected is approximately 1.7\%.
The invariant masses of these seven events
are consistent with $m_{\mathrm{Z}}$.

For the $\bbtt$ selection the ${\cal L}_{\mathrm{\tau\tau}}$ 
cut of the $\qqtt$ selection is
loosened and combined with ${\cal L}_{\mathrm{bb}}$ as follows.
${\cal L}_{\tau\tau}$ and ${\cal L}_{\mathrm{bb}}$ are both
required to be greater than 0.1.
The  $\bbbar\tautau$ probability for the event,
${\cal L}_{\mathrm{bb} \tau\tau}$, is calculated from
Equation~(\ref{eqn:ttprob}) with
${\cal L}_{\tau\tau}$ and  ${\cal L}_{\mathrm{bb}}$ as inputs
and required to
exceed 0.95. 
After the cut on ${\cal L}_{\mathrm{bb} \tau\tau}$,
the remaining cuts of the $\qqtt$\ selection are applied.
None of the $\qqtt$ events are identified as $\bbtt$ candidates
which can be compared with the Standard Model expectation of
$1.41\pm 0.19$.
One additional candidate event is found in the sample with
the relaxed cut on  ${\cal L}_{\mathrm{\tau\tau}}$ 
with $0.35\pm 0.10$ expected.
See Table~\ref{tab:zzxsum} (lines i and j).

\subsection{Selection of {\mybold $\zztoqqnn$}\ events }
\label{ssec:qqnn}
The $\qqnn$ selection is based on
the reconstruction of the $\PZ$ boson
decaying to $\qqbar$ which produces slightly boosted back-to-back jets.  
The
selection uses events
with a two-jet topology
where both jets are contained in the detector.
The beam
energy constraint is used to determine the mass
of the  $\PZ$ boson decaying to $\nunu$.  
The properties of the $\qqbar$ decay and the inferred mass
of the $\nunu$ decay are then used in a likelihood analysis
to separate signal from background.

Two-jet events are selected by
dividing each event into two hemispheres using the 
plane perpendicular to the thrust
axis.
The number of tracks in each hemisphere is required
to be four or more.
The
polar angles  of the energy-momentum vector 
associated with each hemisphere, 
$\theta_{\mathrm{hemi1}}$
and $\theta_{\mathrm{hemi2}}$, are used to define
the quantity 
${\cos\theta_{\mathrm h}} \equiv
       \frac{1}{2}(\cos\theta_{\mathrm{hemi1}}
- \cos\theta_{\mathrm{hemi2}})$. 
Contained events are selected by requiring
$|{\cos\theta_{\mathrm{h}}}| < 0.80$. 
The total energy in the forward detectors and
in the forward region of the electromagnetic
calorimeter ($|\cos\theta|>0.95$) is required to be less than 3\,GeV.
$\PW$~boson decays identified  by the OPAL $\PW$-pair selection are rejected
using the likelihood function
for $\epem \to \qqlnu$ from Reference~\cite{bib:OPAL-WW}
which includes an optimization for each center-of-mass energy.
Only events with $\LWW < 0.5$ are retained.

An important background to our selection is
$\qqbar (n\gamma)$ events with photons that escape 
detection.  We discriminate against these events
by looking for a significant amount of
missing transverse momentum, $\pt$.
In each event, $\pt$ can be resolved into
two components, $\pti$, perpendicular to 
the transverse component of the thrust axis and 
$\ptj$, along the transverse component of the 
thrust axis.
$\pti$ depends primarily on the reconstructed angles of the jets and 
therefore is more precisely measured than $\ptj$ which
depends on the energy balance of the jets.
We approximate $\pti$ as 
$\pti  = \frac{1}{2} \Ebeam \sin\phi \sin\theta_{\mathrm h}$.
Here $\Ebeam = \roots /2$ is the beam energy,
$\phi$ is the acoplanarity calculated 
from the angle between the transverse components of the 
momentum
vectors of the two hemispheres
and  
$\sin\theta_{\mathrm h} = 
\sqrt{ 1 - \cos^2 \theta_{\mathrm h} }$.
The resolution on $\pti$, $\sigma_{\pti}$, was parameterized
as a function of  thrust and ${\cos \theta_{\mathrm h}}$ using
data taken at the $\PZ$ resonance.
We construct the variable
\begin{equation}
R_{\pti} = (\pti - \pti^0)/\sigma_{\pti},
\end{equation}
which is used as input to the likelihood function described below.
Here $\pti^0$ corresponds to the transverse
momentum carried by a
photon with half the beam energy 
which just misses the inner edge of our acceptance 
( $\pti^0 = \Ebeam \sin(32\,\mbox{mrad})/2$).
The likelihood function also uses the related variable
$\cos\theta_{\mathrm{miss}}$, the direction of the
missing momentum in the event, to discriminate
against the $\qqbar (n\gamma)$ events.

In the final selection of events, we use a likelihood function based
on the following five variables:
\begin{enumerate}
\item the normalized 
sum of visible and recoil masses $(\mvis + \mrec)/ \roots $,
\item the difference of visible and recoil masses ($\mvis - \mrec$),
\item
$\log(y_{23})$, where $y_{23}$ is the jet resolution parameter 
that separates the two-jet
topology from the three-jet topology as
calculated from the Durham jet algorithm,
\item $\cos\theta_{\mathrm{miss}}$ and
\item $R_{\pti}$.
\end{enumerate}
The mass variables are useful for reducing background from
$\PW$-pair production and single $\PW$ ($\Wenu$) final states. 
The jet resolution parameter is useful in reducing
the remaining $\qqlnu$ final states.
To improve the performance of the likelihood analysis we
use only events with  $| \mvis - \mrec | < 50$\,GeV,
$(\mvis + \mrec) > 170$\,GeV and $R_{\pti}>1.2$.
Events are then selected using
$\Lqqnn > 0.5$, where $\Lqqnn$ is the likelihood function for
the $\qqnn$ selection.
The likelihood distribution of data and Monte Carlo simulation is shown in
Figure~\ref{fig:zzlikenn}c. 
For the $\bbnn$ selection we require, in addition,
the b-tag variable of 
Reference~\cite{bib:opal-higgs} to be greater than 0.65.

The mean efficiency for the $\qqnn$ selection alone is
$32\%$ and does not have a strong dependence 
on $\roots$.
The efficiency includes corrections to the Monte Carlo
simulation for the effects of backgrounds in the foward detectors
and for imperfect detector modeling.
The modeling systematic
uncertainties are discussed below in Section~\ref{ssec:sys}.  
A total of 60 candidates is observed in the
data to be compared with the Standard Model
expectation of $64.9 \pm  2.9$.
The efficiencies after considering the results of
the b-tagging, as well as the
number of events selected are given in
Table~\ref{tab:zzxsum} (lines~\qqnnlb\ and \bbnnlb).

\subsection{Selection of {\mybold $\zztoqqqq$}\ events }
\label{sec:qqqq}

As in the previous analysis~\cite{bib:zzopal} the hadronic selection
consists of a preselection followed by two different likelihood selections,
one of them aiming at an inclusive selection of fully hadronic 
$\PZ$-pair decays, and the other optimized for selecting final
states containing at least one pair of b quarks.  The main
changes with respect to the previous analysis are that cuts in
the preselection have been modified slightly and that the quantities
used in the $\qqbb$ likelihood have changed.

\subsubsection{Preselection}
The preselection starts from the inclusive multihadron selection
described in Reference~\cite{bib:OPAL-SM172}\@.
      The radiative process $\epem \to \PZ \gamma \to \qqbar \gamma$ 
      is suppressed
      by requiring the effective center-of-mass energy after
      initial-state radiation,
      $\rootsp$, to be larger than 160\,GeV\@.
      $\rootsp$ is  obtained from
      a kinematic fit~\cite{bib:OPAL-SM172} that allows 
      for one or more radiative
      photons in the detector or along the beam pipe.
      The final-state particles are then grouped into jets using
      the Durham algorithm~\cite{bib:durham}.
      A four-jet sample is formed by requiring 
      the jet resolution parameter $y_{34}$ to be at least
      $0.003$ and each jet to contain at least two
      tracks.
      In order to suppress $\PZ^*/\gamma^* \to \qqbar$ background,
      the event shape parameter $C_{\mathrm{par}}$
      \cite{bib:cpar}, which
      is large for spherical events, is required to be greater than 0.27.
      A 4C kinematic
      fit using energy and momentum conservation
      is required to converge.
      A 5C 
      kinematic fit which forces the two jet pairs to have the same mass
      is applied in turn to all three possible
      combinations of the four jets.  This fit is
      required to converge for at least one combination.

The most probable pairing of the jets is determined 
using a likelihood
discriminant which is based on the difference between the two di-jet masses
calculated from the results of a 4C kinematic fit, the di-jet mass
obtained from a 5C kinematic fit, and the $\chi^2$ probability of
the 5C fit.

\subsubsection{Likelihood for the inclusive 
{\boldmath $\PZ\PZ \to \qqbar\qqbar$} event selection}
We use a likelihood selection with six input variables for the
selection of
$\PZ\PZ \to \qqbar\qqbar$ events.
The likelihood selection is optimized separately for each
energy point.
The first variable is the output value of the jet pairing likelihood
described above.
The second variable is determined by
excluding the jet pairing with the largest difference between
the two di-jet masses as obtained from the 4C fit, and then
evaluating the  5C-fit masses of 
the remaining two pairings.
The one which
is closest to the  $\PW$ mass is selected
and
the difference between this 5C-fit mass and the
$\PW$ mass is used in order to discriminate against hadronic  
$\PW$-pair events.
The third variable 
suppresses background events with radiated photons
that shift the value of the mass
obtained from the kinematic fit.
We use the
value of $\rootsp$ from the preselection.
The fourth variable is used
to discriminate against $\PZ^*/\gamma^*$ events.
We use the difference between the largest and
smallest jet energies after the 4C fit.
The final two variables 
are calculated from the momenta of the four jets.
They are
the effective matrix elements for the QCD processes
$\PZ^*/\gamma^\star\to\qqbar{\rm gg}$ and 
$\PZ^*/\gamma^\star\to\qqbar\qqbar$
as defined in Reference~\cite{bib:seymour},
and
the matrix element for the process
${\rm WW}\to\qqbar\qqbar$ from Reference~\cite{bib:excalibur}\@.

The cut on the likelihood function 
has been chosen in order to minimize
the total expected relative error
when including a 10\% relative systematic
uncertainty on the background rate in addition to
the expected statistical error.
The $\qqqq$ likelihood function for the data 
is shown in Figure~\ref{fig:zzlikeqq}a.
A total of 206 events is observed
which can be 
compared to the total Standard Model expectation 
of $201.9 \pm  13.6$ which includes a large background
of $135.6 \pm  13.1$ events.
In Table~\ref{tab:zzxsum}  (lines~\qqqqlb\ and \qqbblb) 
we give the efficiency, background 
and observed number of events.

\subsubsection{Likelihood function for 
{\boldmath $\PZ\PZ \to \qqbar\bbbar$} event selection}

\begin{figure}
\begin{center}
\epsfig{file=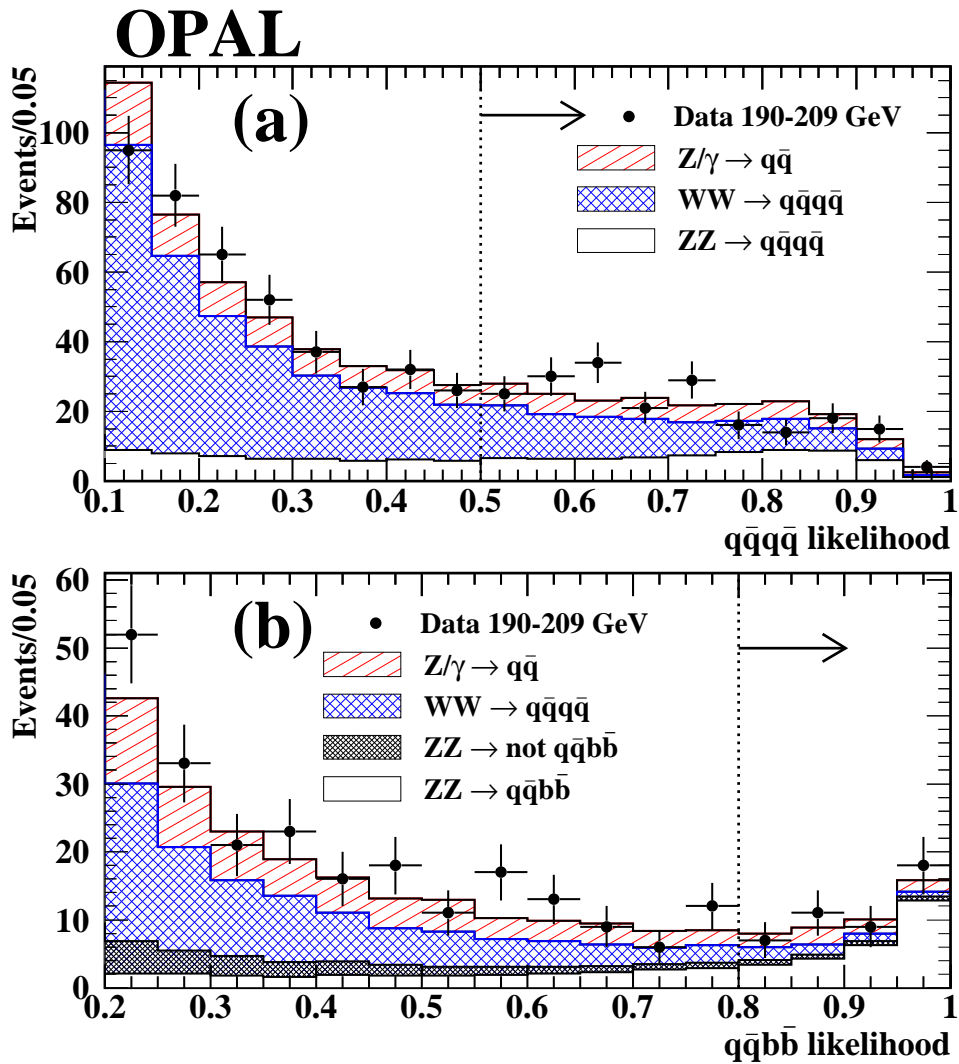,width=0.90\textwidth}
\end{center}
\vspace{-1.0cm}
\caption[$\qqqq$ likelihood ]{
The combined distributions of likelihood discriminants
used for all energies above 190\,GeV.
(a) $\eetozzqqqq$ selection
and 
(b) $\eetozzqqbb$ selection.
The dotted line and arrow show the position of the likelihood cut used
to select ZZ events in each case.
\label{fig:zzlikeqq}
}
\end{figure}

Jets originating from b quarks are selected using the 
b-tagging algorithm~\cite{bib:opal-higgs}.
We evaluate the probability for each of the four jets to originate from
a primary b quark. The two highest probabilities are then
used as input variables
for a likelihood function to select $\PZ\PZ \to \qqbar\bbbar$ events.
In addition
we use the parameters $y_{34}$, $C_{\mathrm{par}}$, the track
multiplicity of the event and the output of the jet
pairing likelihood function. We also use the fit probability of a 6C fit which
forces both masses to be equal to the $\PW$ mass.
Finally we use the variable $|\cos\theta_{\rm bb}-\cos\theta_{\rm qq}|$,
where $\theta_{\rm bb}$ is the opening angle between the most likely
b jets, and $\theta_{\rm qq}$ is the opening angle between the remaining
two jets\footnote{
In contrast to our previous publication~\cite{bib:zzopal},
the probability of the 5C mass fit that constrains one
of the candidate $\PZ$ bosons to the $\PZ$ mass,
as used in the OPAL Higgs~\cite{bib:opal-higgs} analysis,
is not an input to the likelihood function.}.
The $\qqbb$ likelihood function for the combined data taken at
$\roots > 190$\,GeV
is shown in Figure~\ref{fig:zzlikeqq}b.
A total of 45 events is observed
which can be 
compared to the total Standard Model expectation 
of $43.3 \pm  1.8$.
In Table~\ref{tab:zzxsum}  (lines~\qqxblb\ and \qqbblb) 
we give the efficiency, background 
and observed number of events.

{\small
\begin{table}
\begin{center}
\vskip 0.2cm
$\roots = 190 - 209$\,GeV \\
\vskip 0.2cm
{\small
\begin{tabular}{|c|l|ccccccc|}
\hline
&Selection & $\nobs$ & $\smtot$ &$\nsm$ & $\nback$ &$\eff$ & $\br$ & $\lint$\\
&          &         &          &       &          &       &       &
 $(\pb^{-1})$  \\
\hline
 a&$\llllz$&  4&$ 3.55 \pm  0.29$ & $ 2.46 \pm  0.09$ & $ 1.08 \pm  0.27$ & $0.56 \pm 0.02$ &  0.010& 433.6 \\ 
 b&$\eennz$&  2&$ 3.71 \pm  0.35$ & $ 2.43 \pm  0.31$ & $ 1.28 \pm  0.15$ & $0.41 \pm 0.05$ &  0.013& 435.2 \\ 
 c&$\mmnnz$&  0&$ 4.30 \pm  0.39$ & $ 2.39 \pm  0.33$ & $ 1.91 \pm  0.21$ & $0.41 \pm 0.06$ &  0.013& 435.2 \\ 
 d&$\qqeez$& 14&$ 11.6 \pm   0.4$ & $  9.6 \pm   0.3$ & $ 1.98 \pm  0.23$ & $0.62 \pm 0.03$ &  0.037& 424.7 \\ 
 e&$\bbeez$&  3&$ 2.49 \pm  0.19$ & $ 2.20 \pm  0.18$ & $ 0.30 \pm  0.07$ & $0.51 \pm 0.05$ &  0.010& 424.7 \\ 
 f&$\qqmmz$& 15&$ 12.9 \pm   0.4$ & $ 12.1 \pm   0.4$ & $ 0.83 \pm  0.12$ & $0.77 \pm 0.03$ &  0.037& 424.7 \\ 
 g&$\bbmmz$&  7&$ 2.59 \pm  0.15$ & $ 2.43 \pm  0.14$ & $ 0.16 \pm  0.06$ & $0.56 \pm 0.05$ &  0.010& 424.7 \\ 
 h&$\qqttz$&  4&$ 5.35 \pm  0.41$ & $ 4.63 \pm  0.39$ & $ 0.72 \pm  0.12$ & $0.30 \pm 0.03$ &  0.037& 424.7 \\ 
 i&$\bbttz$&  0&$ 1.41 \pm  0.19$ & $ 1.19 \pm  0.18$ & $ 0.21 \pm  0.06$ & $0.28 \pm 0.04$ &  0.010& 424.7 \\ 
 j&$\xbttz$&  1&$ 0.35 \pm  0.10$ & $ 0.29 \pm  0.10$ & $ 0.07 \pm  0.03$ & $0.07 \pm 0.02$ &  0.010& 424.7 \\ 
 k&$\qqnnz$& 51&$ 56.4 \pm   2.8$ & $ 30.3 \pm   2.5$ & $ 26.1 \pm   1.2$ & $0.33 \pm 0.03$ &  0.219& 422.1 \\ 
 l&$\bbnnz$&  9&$ 8.45 \pm  0.74$ & $ 7.23 \pm  0.71$ & $ 1.22 \pm  0.22$ & $0.28 \pm 0.03$ &  0.061& 422.1 \\ 
 m&$\qqqqz$&185&$180.5 \pm  13.5$ & $ 50.2 \pm   3.1$ & $130.3 \pm  13.1$ & $0.39 \pm 0.03$ &  0.300& 432.3 \\ 
 n&$\qqxbz$& 24&$ 21.9 \pm   1.4$ & $ 13.7 \pm   0.8$ & $ 8.14 \pm  1.14$ & $0.17 \pm 0.01$ &  0.189& 432.3 \\ 
 o&$\qqbbz$& 21&$ 21.4 \pm   1.2$ & $ 16.1 \pm   0.9$ & $ 5.30 \pm  0.74$ & $0.20 \pm 0.02$ &  0.189& 432.3 \\ 

\hline
\end{tabular}     
}
\end{center}
\caption[summary]{ 
Observed number of 
events, $\nobs$, the total
Standard Model expectation, $\smtot$,
the expected number of $\PZ$-pairs, $\nsm$,
background expectation, $\nback$, and
efficiencies, $\eff$, for the combined data sample. 
$\br$ is the product branching
ratio for the final state
which is calculated directly from 
$\PZ$ resonance data~\cite{bib:pdg}.
An overbar
is used to indicate that events
from a particular selection are rejected.
Note that the efficiencies 
for selections with b-tags are given relative to 
the fraction of hadronic final states which
contain a $\PZ$ boson decaying to $\bbbar$.  
For selections of events with hadronic final states, but
without b-tags, the efficiencies are relative to
those expected hadronic final states which do not include
a $\PZ$ boson decaying to $\bbbar$.
The errors on all quantities
include contributions from the
systematic uncertainties described in Section~\ref{ssec:sys}.
\label{tab:zzxsum}
}
\end{table}
}

\subsection{Systematic uncertainties}
\label{ssec:sys}

Systematic uncertainties have only a modest effect on our final 
result because of the large statistical
error associated with the small 
number of $\PZ$-pair events produced at LEP\,2.

Detector systematic 
uncertainties for the $\qqll$ and $\llll$ selections 
without $\tau$-pairs
in the final state are small because of the good separation
of signal and background.
In the $\qqll$ channels possible mismodeling of the
detector resolution is accounted for
by smearing polar and azimuthal jet 
angles with a Gaussian width of 1$^\circ$ and the energies by
5\%\@. This gives systematic shifts smaller than 5\%\@.
These shifts are included in the systematic uncertainties on the
efficiencies.
In the $\qqtt$ selection, the additional systematic
uncertainties in the  efficiencies are determined by
overlaying hadronic and tau decays taken from $\PZ$
resonance data giving the dominant contribution to the 
total systematic uncertainty of 6.2\%.
In the $\llll$ final state the
largest effect (3\%) is
from the modeling of the multiplicity requirement which 
is important for final states containing $\tau$-pairs.

In the final states with neutrinos,  $\llnn$ and $\qqnn$, 
tight cuts are needed to separate signal and background.
In these selections 
detector effects can best be studied by comparing calibration
data taken at the $\PZ$ resonance with a simulation of
the same process.
In these
cases, we add additional smearing to the total
energy and momentum
of the simulated events to match data and simulation
on the $\PZ$-resonance. 
We then apply the same smearing to the signal and background
Monte Carlo simulations for the $\PZ$-pair analyses. 
The reported efficiencies and backgrounds are
accordingly corrected.
The full difference is used as the systematic
uncertainty in these cases.  
These differences give relative systematic
uncertainties on the efficiency of 2.5\%
for the $\eenn$
final state, 5.2\%  for the $\mmnn$ final state and
3.8\% for $\qqnn$ final state.

In the $\qqqq$ inclusive analysis the sensitivity to
the detector description of jet eneriges is much
reduced by the use of kinematic fits.  
In this case the systematic
uncertainties are determined by smearing the jet energies
by an additional 5\% and the jet directions by 1$^\circ$
leading to a relative detector systematic uncertainty of 6\%.

Another important detector effect comes from the simulation
of the variables used by the OPAL b-tag which is discussed in
Reference~\cite{bib:opal-higgs}.  We allow for a common
5\% uncertainty on the efficiency of the b-tag, consistent with our
studies on $\PZ$ resonance data and Monte Carlo simulation.

In each channel the signal and background Monte Carlo 
generators have been compared against alternative
generators.  
In almost all cases
the observed differences are consistent within the finite
Monte Carlo statistics and the systematic uncertainty has been
assigned accordingly.

The largest contribution to the systematic uncertainty from
the  model dependence 
of the background prediction is for the $\qqqq$ and $\qqbb$
final states.
This uncertainty has been estimated by
comparing the predictions of KORALW and grc4f for
the $\PW$-pair background, 
and the prediction of PYTHIA and KK2f for the 
background from hadronic 2-fermion production. 
The resulting uncertainty
is 10\% on the background for the inclusive selection and 20\% for 
the background of the $\qqbb$ selection~\cite{bib:opal-higgs}.
The 10\% uncertainty has been taken to be fully correlated between
energy points and between the $\qqqq$ and $\qqbb$ selections.
The  correlated 
error on the $\qqbb$ selection from uncertainties in
the background due to b-tagging, has been absorbed
into the overall b-tagging efficiency uncertainty.

In the $\llll$ channel, which has
a large background from two-photon events,
we have compared the number of selected events
at an early stage of the analysis with the Monte Carlo simulation
and based our background systematic uncertainty on the level of
agreement.  This results in 20\%\@ systematic uncertainty
on the background.

The efficiencies and backgrounds given for each energy point
(see the Appendix and the summary 
Table~\ref{tab:zzxsum}) include the systematic uncertainties described in
this section as well as errors from finite Monte Carlo statistics.
Given the large statistical error from the small $\PZ$-pair 
cross section we have employed the following conservative scheme
to account for possible 
correlations among the systematic uncertainties:
\begin{enumerate} 
\item The common correlated error from detector systematics,
hadronization and kinematic distributions of the 
$\PZ$-pair events is taken to be 3\%.  
This is based primarily on the studies of the smearing
of reconstructed jet and lepton directions.  
\item The common systematic
uncertainty for any channel including a b-tag is taken to be 5\%.  
\item  As described above an
uncertainty of 10\% of the backgrounds is taken to be fully correlated 
among all $\qqqq$ and $\qqbb$ channels.  
\item Based on generator level comparisons of kinematic distributions
we conservatively assume  a 2\%\@  correlated systematic uncertainty for
the modeling of any changes to  the physics description of the $\PZ$ process
when new physics is switched on.  This is in addition
to an overall uncertainty of 2\%\@ for the Standard Model 
total cross section prediction of the ZZTO calculation.
\end{enumerate}

\section{Cross section and branching ratios}                    \label{sec:xsec}
At each center-of-mass energy point,
information from all of the analyses is combined
using a maximum likelihood fit
to determine the production cross section for
$\eetozz$.
The information which was used in the fits, as
well as the Standard Model prediction for $\PZ$-pair
production, 
is summarized in Table~\ref{tab:zzxsum}.
The details for each energy point
are given in the Appendix.
For each channel the tables give the number of events observed, 
$\nobs$, the Standard Model prediction for
all events, $\smtot$, the expected
signal, $\nsm$, 
the expected background, $\nback$, 
the efficiency $\eff$, 
and the integrated luminosity, $\lint$.
$\br$ is the branching ratio of $\PZ$-pairs to 
the given final state, calculated from $\PZ$
resonance data~\cite{bib:pdg}.
In the table we give the overlap between 
the b-tag and non-b-tag analyses.
Possible overlap between $\qqqq$ and $\qqll$ has been studied,
and found to be
an order of magnitude smaller than the overlap of 
$\qqqq$ and $\qqbb$ and has therefore been ignored.

\begin{figure}
\begin{center}
    \mbox{\epsfxsize16cm\epsffile{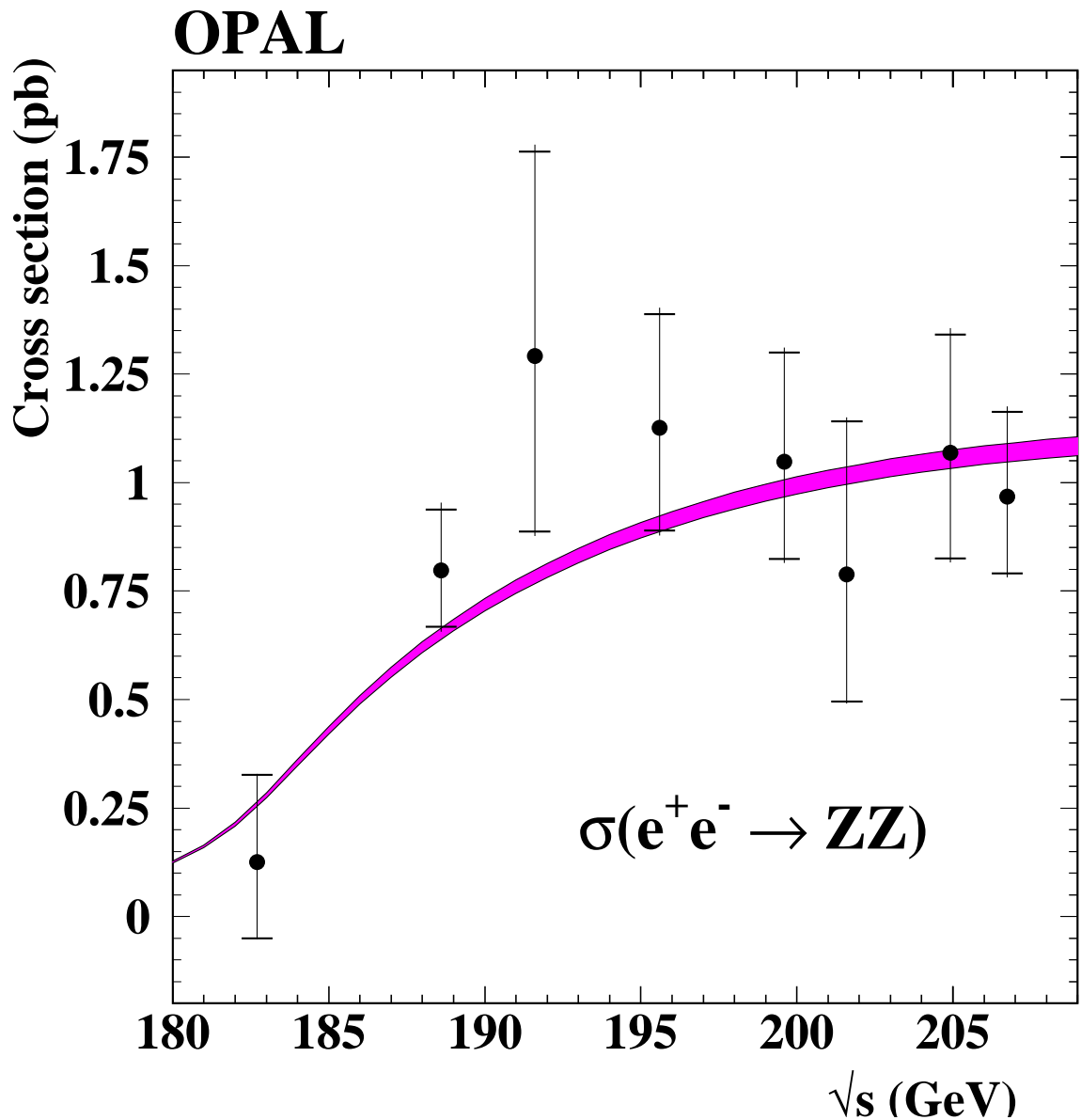}}
\end{center}
\caption[ZZ cross section]{
\label{fig:zzxsec}
The OPAL measurements
of the NC2 $\PZ$-pair production cross section.  
The middle of the shaded band shows the Standard Model prediction of 
the ZZTO calculation.
The
band indicates a theoretical  uncertainty
of $\pm 2$\%.
The error bars on the data show the combined statistical 
and systematic uncertainty,
the horizontal marks indicate the extent of the
statistical errors.
Data from  
$\roots$ near 183\,GeV and 189\,GeV are from Reference~\cite{bib:zzopal}.
}
\end{figure}

\begin{figure}
\begin{center}
\epsfig{file=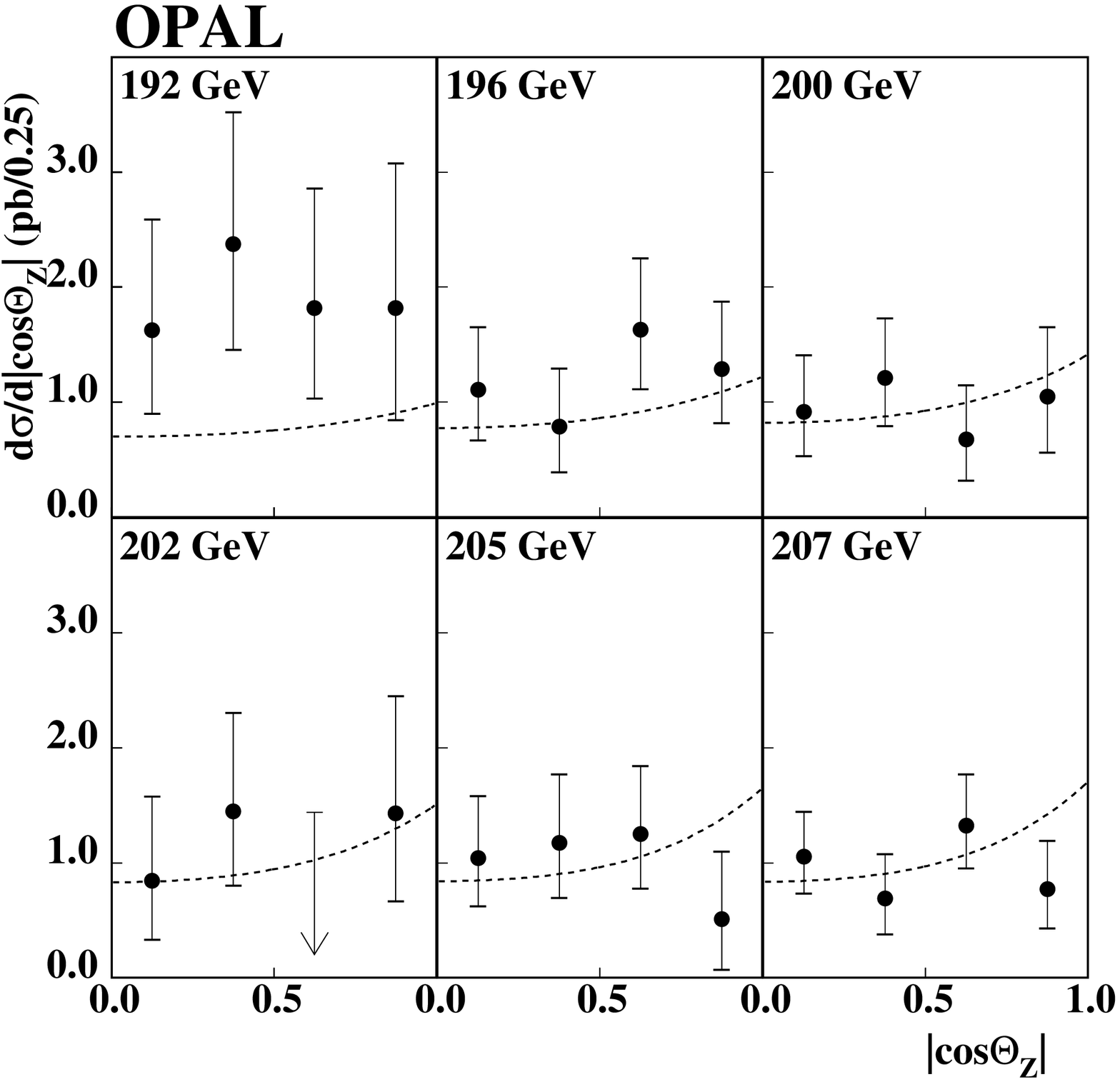,width=0.95\textwidth}
\end{center}
\caption[ZZ cross section]{
\label{fig:zz_cos}
The OPAL measurements
of the NC2 $\PZ$-pair 
differential cross section 
$\mathrm{d} \sigma_{\mathrm{ZZ}}/ \mathrm{d} |\cos \theta_{\mathrm{Z}} |$.  
The curves show the prediction
of the YFSZZ Monte Carlo for the differential cross section,
which has been 
normalized to agree with the total cross section of 
the ZZTO calculation.
The error bars show the combined statistical 
and systematic uncertainty.  The arrow shows the 95\% confidence
level upper limit for
the one  case where the minimum of the negative log likelihood
function was at zero.
}
\end{figure}

The cross section at each energy is determined with
a maximum likelihood fit using  Poisson probability
densities convolved with Gaussians to describe the
uncertainties on efficiencies and backgrounds.  
The expected number of events in each channel, $\nexp$,
as function of the $\PZ$-pair cross section, $\xsec$, 
is given by
\begin{equation}
\nexp  =  \xsec \lint \eff \br  + \nback.
\end{equation} 
The efficiencies, $\eff$,
include the effects of off-shell $\PZ$ bosons that 
are produced
outside of our kinematic acceptance.
The correlated systematic uncertainties described in the 
last section are implemented by introducing additional
parameters which are constrained with Gaussian
probability densities given by the size of the systematic 
uncertainty.
Our main results, 
the NC2 $\PZ$-pair cross sections obtained from the
fits, are
$$
\begin{array}{lcll}
\xsec  (192 \ \mathrm{GeV})   & = & \xscc & \mathrm{pb} \\
&&& \\
\xsec  (196 \ \mathrm{GeV})  & = & \xscd & \mathrm{pb} \\
&&& \\
\xsec  (200 \ \mathrm{GeV})   & = & \xsce & \mathrm{pb} \\
&&& \\
\xsec  (202 \ \mathrm{GeV})  & = & \xscf & \mathrm{pb} \\
&&& \\
\xsec  (205 \ \mathrm{GeV})  & = & \xscg & \mathrm{pb} \\
&&& \\
\xsec  (207 \ \mathrm{GeV})  & = & \xsch & \mathrm{pb} .\\
\end{array}
$$
The first error is statistical and the second error is systematic.
The systematic errors are obtained from the
quadrature difference of the errors when running the fit
with and without systematic uncertainties. 
The comparison of these measurements
with the ZZTO prediction 
is shown in Figure~\ref{fig:zzxsec}.  
The results are consistent with the Standard Model prediction.

We have also analyzed our data in four bins of 
$|\cos \theta_{\mathrm{Z}}|$ where $\theta_Z$ is the polar
angle of the $\PZ$ bosons. 
Except for channels with one $\PZ$ boson decaying to
neutrinos, the value of
$|\cos \theta_{\mathrm{Z}}|$ 
is determined from kinematic fits which 
assume no initial-state radiation.  In the $\llnn$ and $\qqnn$
channels, the direction is determined from the
reconstructed direction of the visible $\PZ$. 
The comparison of the expected and observed differential cross sections
is shown in Figure~\ref{fig:zz_cos}.  The data in Figure~\ref{fig:zz_cos}
have been corrected for small amounts of bin migration assuming
the Standard Model prediction.

\begin{figure}
\begin{center}
\epsfig{file=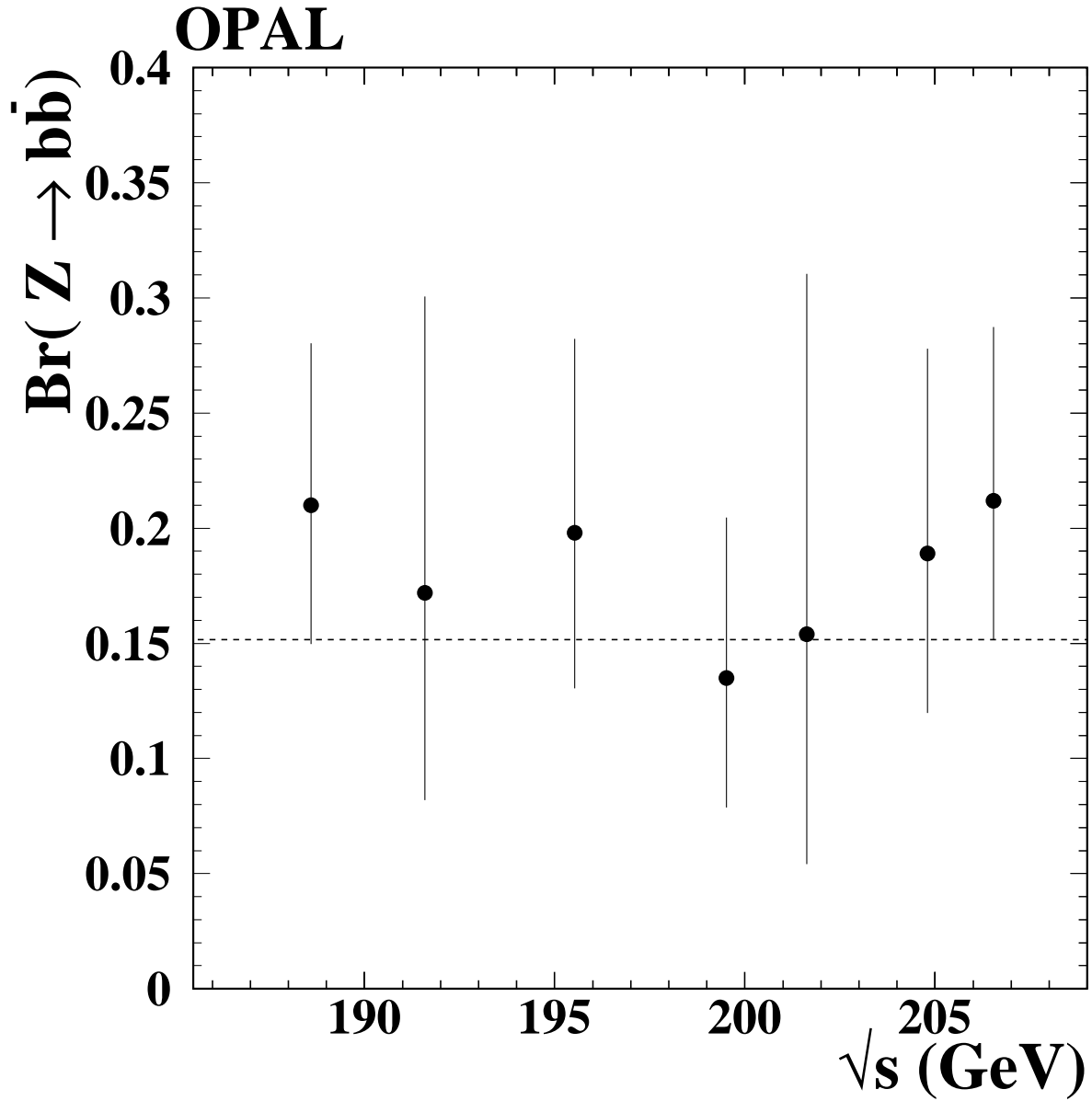,width=0.95\textwidth}
\end{center}
\caption[Z branching ratio]{
\label{fig:zzbr}
Determination
of Br$( \PZ \to \bbbar)$ from $\PZ\PZ$ events.  
The line shows the measured value from LEP\,1 
data.
Only statistical errors are shown.  The data
have a common systematic uncertainty which amounts
to approximately 5\% of the expected branching ratio.
The 189\,GeV point is from
Reference~\cite{bib:zzopal};
at 183\,GeV the \ZZ\ event samples were too small to allow
a meaningful measurement to be obtained.
}
\end{figure}

In order to check the consistency of our result
with the Standard Model, we have performed a maximum likelihood
fit in which the Standard Model ZZTO prediction is scaled by an 
overall factor $R$.  
The fit is based on individual measurements for each channel
given in the Appendix in 
Tables~\ref{tab:zzxsec192} to \ref{tab:zzxsec207} as well as 
the data presented in Reference~\cite{bib:zzopal}.
The treatment of the correlated systematic
uncertainties is outlined in Section~\ref{ssec:sys}.
The fit yields
$$
R = 1.06 ^{+0.11}_{-0.10}
$$
where the error includes the correlated experimental systematic
uncertainty (3\%) and the theoretical uncertainty on the ZZTO
prediction (2\%).
In a separate fit, we
allowed the
branching ratio  of $\PZ$ bosons to b quarks, 
BR$(\PZ \to \bbbar)$, to be a
free parameter in the fits at each of the
energy points.  The results are shown in Figure~\ref{fig:zzbr}.
The BR$(\PZ \to \bbbar)$ values tend to lie above the measured value
from LEP\,1 data; combining all energies, the average value is
0.196$\pm$0.032, which is 1.3 standard deviations above the LEP\,1
measurement of 0.1514$\pm$0.0005\cite{bib:pdg}.

\section{Limits on anomalous triple gauge couplings}      \label{sec:ac}

Limits on anomalous triple gauge couplings were set using
the measured cross sections and  kinematic information
from  an optimal observable (OO) 
method for the $\qqqq$, $\qqbb$  and $\qqll$ selections.
In this study we vary the real 
part of the $\PZ\PZ\PZ$ and $\PZ\PZ\gamma$ 
anomalous couplings parameterized by
$\ZZZ{4}$,  $\ZZZ{5}$, 
$\ZZG{4}$ and $\ZZG{5}$ as
defined in Reference~\cite{bib:hagiwara} and
implemented in the YFSZZ Monte Carlo.
In most cases, the real parts of each coupling 
were varied separately with all others
fixed to zero.

The OO analysis used here is described in detail
in Reference~\cite{bib:markus}.  Since the effective
Lagrangian used to describe the anomalous couplings is linear in
the couplings, the resulting differential
cross section is parabolic and can be parameterized,
for a single non-zero coupling $\alpha_i$, as
\begin{equation}
 \frac{\mathrm{d} \sigma} {\mathrm{d} \Omega} = S^{(0)} (\Omega)                +    
                              \alpha_i S^{(1)}_i (\Omega)     +
                              \alpha_i^2 S^{(2)}_i (\Omega)
\label{eq:sorder}
\end{equation}
where $\Omega$ is the phase-space point based on the
four-momenta of the four out-going fermions from the
$\PZ$ decays.  The optimal observables are defined as
\begin{equation}
\begin{array}{lcl}
{\cal O}_1^i & \equiv & S^{(1)}_i (\Omega) /  S^{(0)} (\Omega) \\
{\cal O}_2^i & \equiv & S^{(2)}_i (\Omega) /  S^{(0)} (\Omega). \\
\end{array}
\end{equation}

To reduce the dependence of our analysis on the tails of
the ${\cal O}_1^i$ and ${\cal O}_2^i$ distribution, 
extreme values of ${\cal O}_1^i$ and  ${\cal O}_2^i$
were rejected when calculating the mean values.
These cuts were chosen based on the expected distribution
of ${\cal O}_1^i$ and ${\cal O}_2^i$ calculated 
from Monte Carlo and typically reject a few percent of 
the expected events.

The expected 
average values of the first order optimal observable, 
$\langle{\cal O}_1\rangle$,
and the second order optimal observable,
$\langle{\cal O}_2 \rangle$,
were calculated using the YFSZZ matrix element 
to reweight accepted events from the
signal Monte Carlo.
The same cuts on extreme values of
${\cal O}_1^i$ and  ${\cal O}_2^i$ were
used in data and Monte Carlo simulation. 
The parameterization used was of the form
\begin{equation}
\begin{array}{lcl}
 \langle{\cal O}_1\rangle & = & \frac{\mbox{$ p_0 + p_1 \alpha + p_2 \alpha^2$}}
                                     {\mbox{$ d_0 + d_1 \alpha + d_2 \alpha^2$}} \\
&&\\
 \langle{\cal O}_2\rangle & = & \frac{\mbox{$q_0 + q_1 \alpha + q_2 \alpha^2$}}
                                     {\mbox{$d_0 + d_1 \alpha + d_2 \alpha^2$}} \\
\end{array}
\label{eq:poly}
\end{equation}
where $\alpha$ is the value of the anomalous coupling and the coefficients
are determined separately for each selection, energy and type of anomalous
coupling.
Note that the mean values of the optimal observables are normalized to
the observed number of events and do not depend on the cross section
which is considered separately below.  This can be seen in 
Equation~(\ref{eq:poly}) by
noting that the polynomial $d_0 + d_1 \alpha + d_2 \alpha^2$ 
parameterizes the
change in cross section with anomalous coupling.
The average value of the optimal observable for the
exclusive $\qqee$, $\qqmm$ and three categories
of $\qqqq$ events ( $\qqqqz$, $\qqbbz$ and  $\qqxbz$ ) were 
calculated from the data at each energy between 190\,GeV and 207\,GeV.
The statistical uncertainties on these values were parameterized
with  a covariance matrix determined 
from high statistics Standard Model signal and background
Monte Carlo simulations.  The covariance matrix was
then scaled to match the number of events observed in the
data.

\begin{figure}
\begin{center}
\epsfig{file=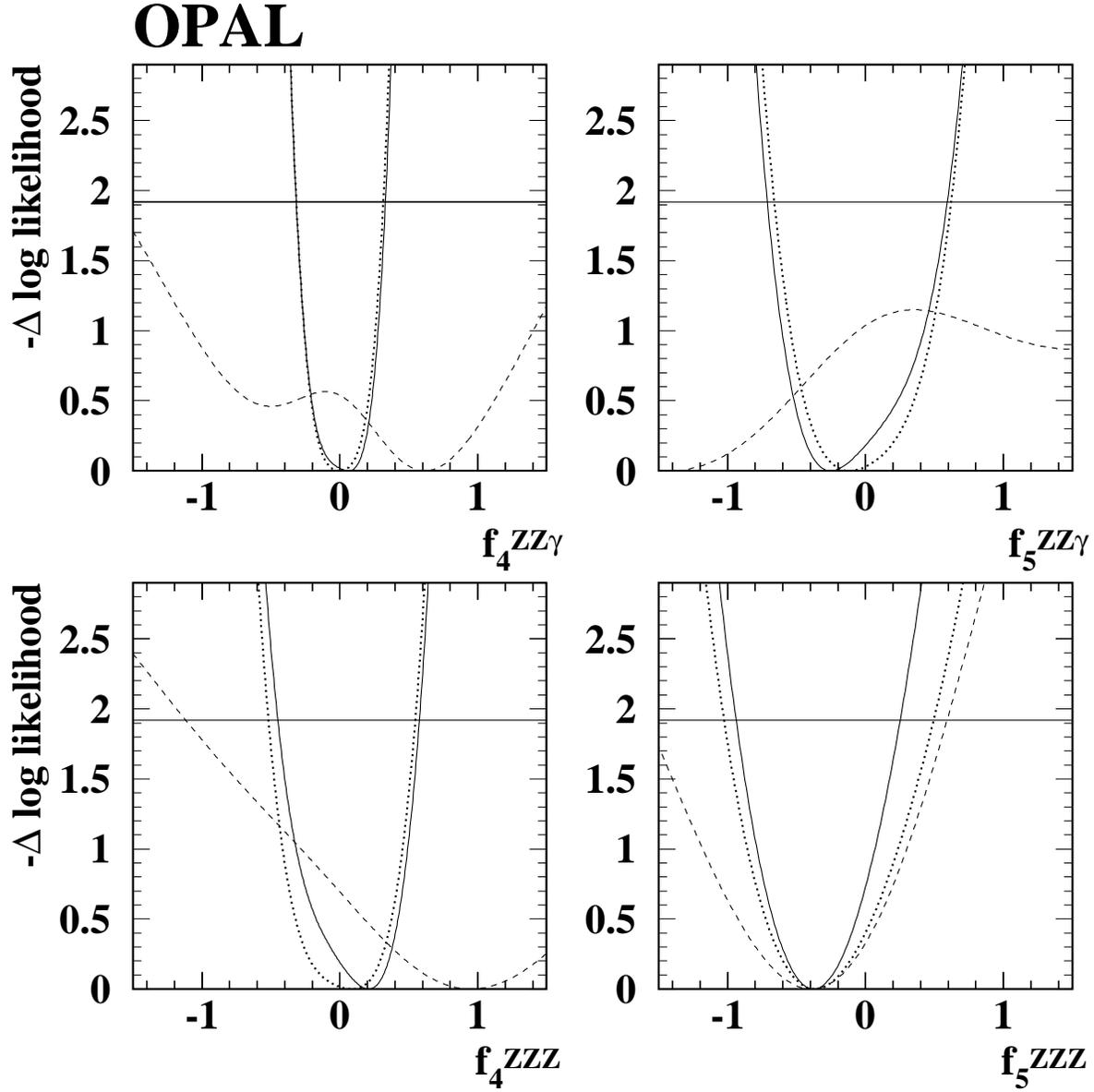,width=0.95\textwidth}
\end{center}
\caption[ZZ Likelihood Curves]{
\label{fig:zzanlike}
The negative log likelihood as a function of the real part
of the four anomalous couplings:
dotted, constraint from cross-section information;
dashed, constraint from optimal observables analysis; and
solid, sum.
The 95\% confidence level corresponds to a change
in log likelihood of 1.92 from the minimum value.
These results are dominated by the $190 - 209$\,GeV data, 
but also include the OPAL data from
183\,GeV and 189\,GeV presented in Reference~\cite{bib:zzopal}.
}
\end{figure}

The systematic uncertainties were taken into account by varying
the modeling of the Standard Model signal and the backgrounds.
Any deviations were interpreted as a systematic error and
included in the covariance matrix.
The error associated with the physics simulation of the signal was
determined by comparing the values of 
$\langle{\cal O}_1\rangle$ and $\langle{\cal O}_2 \rangle$
determined
from reweighting the YFSZZ and grc4f Monte Carlo event samples.  
The error in the background determination was
evaluated by using the alternative background simulations described in
Section~\ref{ssec:sys}.
These differences were taken
to be fully correlated
among energies for a given final state. 
The systematic
uncertainties associated with the accuracy of the event reconstruction
were evaluated by applying additional smearing to the energies
and angles of the reconstructed jets and leptons.  
The resulting contributions to the covariance matrix
were taken to be fully correlated among all five
final state selections. 
The resulting $\chi^2$ functions, converted to 
likelihood curves, are shown as dashed lines in Figure~\ref{fig:zzanlike}.  

For those channels
used in the OO analysis, we use in addition only the cross section
integrated over $|\cos\theta_{\mathrm{Z}}|$.
For channels that are not used in the OO analysis,
the  cross section in four bins of $| \cos \theta_{\mathrm{Z}}|$
is used.
The ZZTO calculation is used for the prediction of the
Standard Model integrated cross section.  The change in 
the $\PZ$-pair cross section as a
function of $|\cos\theta_{\mathrm{Z}}|$
and the values of the anomalous couplings is parameterized 
using the YFSZZ Monte Carlo.
In addition,
the selection efficiencies for all
final states are parameterized
as a function of the couplings and
for each bin in $| \cos \theta_{\mathrm{Z}} |$.
An uncertainty of 10\%, dominated by Monte Carlo
statistical errors, is assigned to the
correction we applied to all of these efficiencies.
The resulting likelihood curves based on
cross-section measurements  are shown as dotted lines in 
Figure~\ref{fig:zzanlike}.  
These cross-section results include 
data at 183\,GeV and 189\,GeV
from Reference~\cite{bib:zzopal}.

Combining the likelihood associated with the  $\chi^2$
fit  and the likelihood curve from the cross-section fit,
the 95\% confidence level (C.L.) limits on the anomalous 
couplings were obtained and
are given in Table~\ref{tab:anlim}.
The combined likelihood
curves are shown as  solid lines in Figure~\ref{fig:zzanlike}.

\begin{figure}
\begin{center}

\epsfig{file=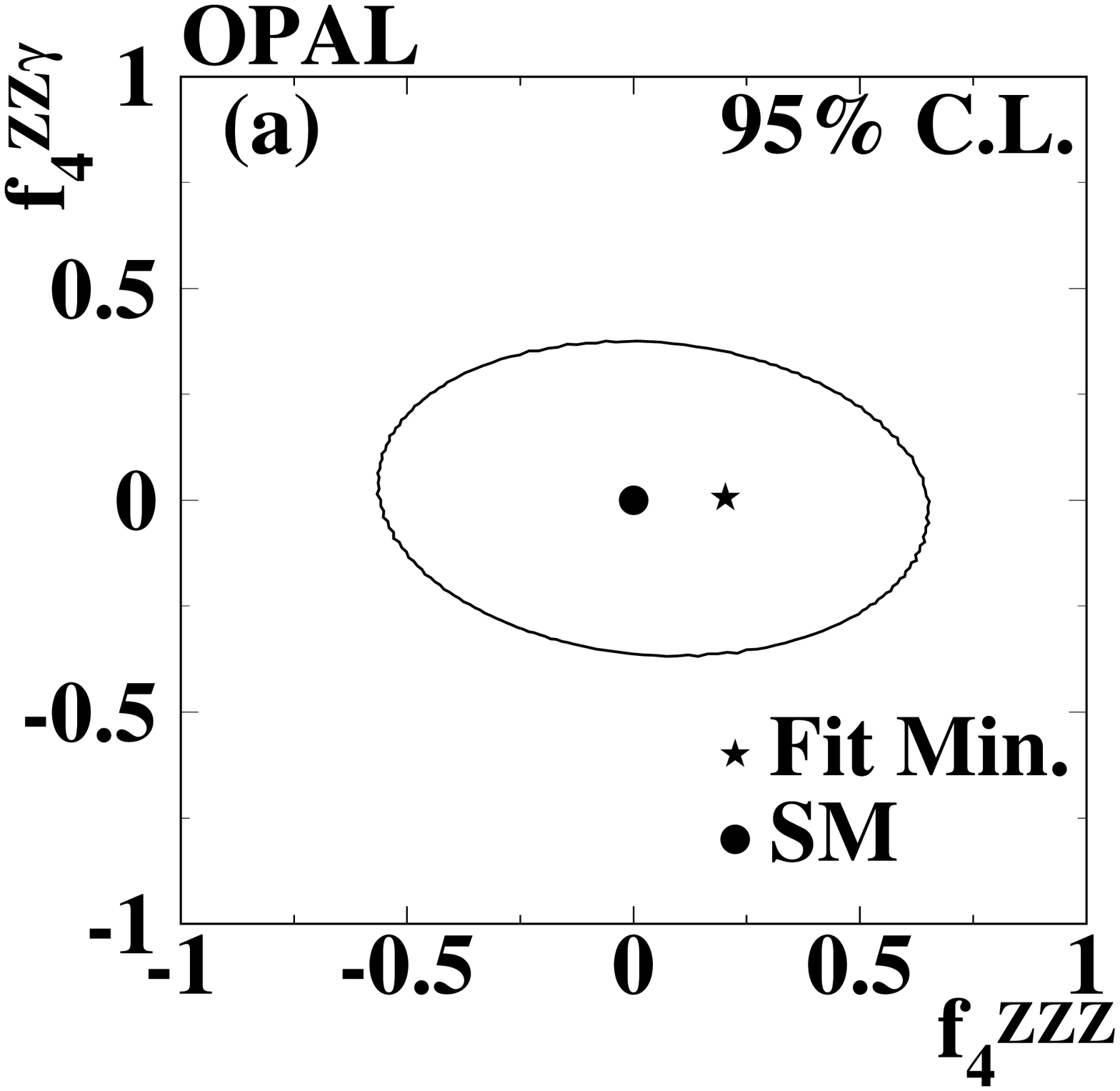,width=0.45\textwidth}\epsfig{file=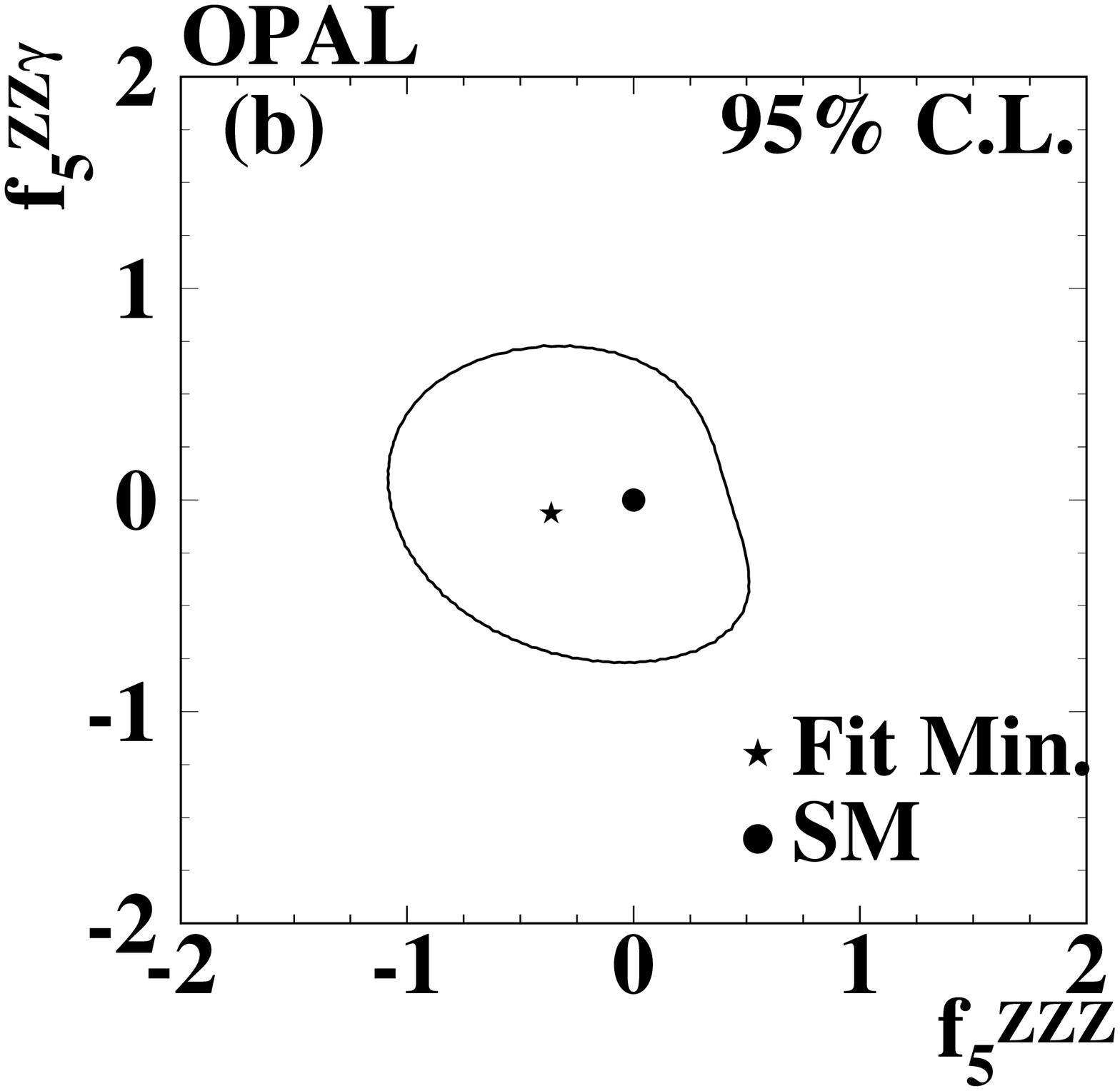,width=0.45\textwidth}
\end{center}
\caption[f4 and f5 planes]{
\label{fig:f4f5}
The 95\% confidence level contours, corresponding to a 
change of 3.0 in the log likelihood from the minimum, for  (a)
the $\ZZG{4}-\ZZZ{4}$ plane and (b)
the $\ZZG{5}-\ZZZ{5}$ plane.
These results include all channels
and all energies,  
including OPAL data from
183\,GeV and 189\,GeV presented in Reference~\cite{bib:zzopal}.
The star shows the location of the minimum, the solid dot the Standard
Model.
}
\end{figure}

Equation~(\ref{eq:sorder}) can easily be extended to the case
of two non-zero couplings and the constraints can be derived
for pairs of couplings~\cite{bib:markus}. 
Note that for the two-dimensional
OO fits, the event sample is slightly different 
from that used in
the one-dimensional
OO fits, as cuts are placed on the values of optimal 
observables for more than one coupling.
The constraint from the expected cross sections can also be
calculated for two non-zero couplings.  
The resulting 95\% confidence level contours in the
$f_4$ and $f_5$ plane, 
including both OO and cross-section constraints, 
are shown in Figure~\ref{fig:f4f5}.
The 95\% confidence level corresponds to a change
in log likelihood of 3.0 from the minimum value.

\begin{table}
\begin{center}
\begin{tabular}{|c|c|c|}
\hline
Coupling & 95\% C.L. lower limit &  95\% C.L. upper limit \\
\hline
 $ \ZZZ{4} $ & $ -0.45 $ & $   0.58$\\
 $ \ZZZ{5} $ & $ -0.94 $ & $   0.25$\\
 $ \ZZG{4} $ & $ -0.32 $ & $   0.33$\\
 $ \ZZG{5} $ & $ -0.71 $ & $   0.59$\\
\hline
\end{tabular}
\end{center}
\caption{
\label{tab:anlim} The 95\% confidence level
limits on possible anomalous triple gauge couplings.
These results include all channels
and all energies,  
including the OPAL data from
183\,GeV and 189\,GeV presented in Reference~\cite{bib:zzopal}.
}
\end{table}

\section{Limits on low scale gravity theories with large extra dimensions} 
     \label{sec:ed}

We have also examined the possible effects of low scale gravity (LSG)
theories with large extra dimensions~\cite{bib:add} 
on the $\PZ$-pair cross section.  
In the LSG theories considered here, gravity is allowed to propagate
in $D = 4+n$ dimensions, while all other particles are confined to
four dimensional space.  Newtonian gravity in three spatial
dimensions holds if
\begin{equation}
M^2_{\mathrm{Planck}} \propto R^n M_D^{(n+2) }
\end{equation}
where $M_{\mathrm{Planck}}$ is the Planck scale in the usual four dimensions,
$M_D$ is the Planck scale in the D-dimensional space,
and $R$ is the compactification radius of the 
$n$ extra dimensions.  
The case of $n=1$ is excluded by cosmological observations.  
For the case of $n=2$, severe constraints are imposed by 
studies of the gravitational interaction on sub-millimeter 
distance scales\cite{bib:long,bib:adelberger}.

Because gravity can propagate in the extra dimensions,
the amplitude for $\PZ$-pair production
has leading order contributions from the
$s$-channel exchange of Kaluza-Klein graviton
states.
The effective theory  at LEP\,2 energies is sensitive to the
mass scale $M_s$ that is used to regulate ultraviolet divergences.
$M_s$ is expected  to be of the same order of magnitude as $M_D$. 
Using this mass scale, $M_s$, a contribution to the 
Born level amplitude that is proportional to $\lambda/M_s^4$ 
is obtained.  Here $\lambda$ is an effective coupling.
In this paper we use the definition of $M_s$ given in
Reference~\cite{bib:joanne} for our main result.

In order to search for deviations from the Standard Model compatible
with LSG theories,  the differential $\PZ$-pair
cross section is written as 
\begin{equation}
\frac{ \mathrm{d} \sigma_{\mathrm{ZZ}}     }
      { \mathrm{d} \cos \theta_{\mathrm{Z}} }      = \\
\frac{\alpha \beta}{\pi s} \sum_{\kappa, \epsilon_+, \epsilon_-} 
C^2_{\Gamma_{\mathrm{Z}}}(s, \cos \theta) 
\left|  {C}^{\mathrm{SM}}_{\mathrm{ISR}} (s)
{\cal M}^{\mathrm{SM}}_{\mathrm{born}}(\kappa,\epsilon_+,\epsilon_-,s,t)    +  
  {C}^{\mathrm{grav}}_{\mathrm{ISR}} (s) 
{\cal M}_{\mathrm{gravity}} (\kappa,\epsilon_+,\epsilon_-,s,t) \right|^2
\label{eq:desh}
\end{equation}
where $\alpha$ is the fine structure constant,
$s$ and $t$ are the Mandelstam variables, $\kappa$ is the electron helicity, 
$\epsilon_+$ and $\epsilon_-$ are the polarizations of the $\PZ$ boson, $\beta$ is
the $\PZ$ velocity, and 
${\cal M}^{\mathrm{SM}}_{\mathrm{born}}(\kappa,\epsilon_+,\epsilon_-,s,t)$
and ${\cal M}_{\mathrm{gravity}}(\kappa,\epsilon_+,\epsilon_-,s,t)$ are
the Born level SM and gravity matrix elements~\cite{bib:desh}.  
Here ${\cal M}_{\mathrm{gravity}}(\kappa,\epsilon_+,\epsilon_-,s,t)$
is proportional to $\frac{\lambda}{M_s^4}$.
The
factor
$C_{\Gamma_{\mathrm{Z}}}(s, \cos \theta )$ 
corrects for the effects of the finite width 
of the $\PZ$. The factors $C^{\mathrm{SM}}_{\mathrm{ISR}}(s)$
and  $C^{\mathrm{grav}}_{\mathrm{ISR}}(s)$  correct for 
initial-state radiation.

The factors $C^{\mathrm{SM}}_{\mathrm{ISR}}$(s), 
and  $C^{\mathrm{grav}}_{\mathrm{ISR}}(s)$ are given by
\begin{equation}
C = 
\sqrt{\frac{\sigma^{\mathrm{YFSZZ}}_{\mathrm{ISR}}  }
           {\sigma^{\mathrm{YFSZZ}}_{\mathrm{Born}} } }
\end{equation}
where $\sigma^{\mathrm{YFSZZ}}_{\mathrm{ISR}}$ is the YFSZZ cross section with
ISR and $\sigma^{\mathrm{YFSZZ}}_{\mathrm{Born}}$ is the Born level 
YFSZZ cross section.
The ISR correction for Standard Model $\PZ$ production,
$C^{\mathrm{SM}}_{\mathrm{ISR}}$(s),
is estimated by running YFSZZ with
all anomalous couplings off.  For the case of gravity, 
$C^{\mathrm{grav}}_{\mathrm{ISR}}(s)$ is estimated
by using a choice 
of anomalous couplings which causes the s-channel to dominate.
These factors range from 0.84 at 183\,GeV to 
0.93 at 207\,GeV.   $C^{\mathrm{SM}}_{\mathrm{ISR}}(s)$ 
and  $C^{\mathrm{grav}}_{\mathrm{ISR}}(s)$ differ from each other
by no more than 5\%.

The effect of the finite $\PZ$ width depends on $|\cos \theta_{\mathrm{Z}}|$ and is
obtained from
\begin{equation}
C_{\Gamma_{\mathrm{Z}}}(s, \cos \theta ) = 
\sqrt{\frac{\sigma^{\mathrm{YFSZZ}}_{|\cos\theta_{\mathrm{Z}}|}}
           {\sigma^{\mathrm{\Gamma_Z = 0}}_{|\cos\theta_{\mathrm{Z}}| }} }
\end{equation}
where $\sigma^{\mathrm{\Gamma_Z = 0}}$ is the prediction of 
Equation~(\ref{eq:desh}) with
$C_{\Gamma_{\mathrm{Z}}}(s, \cos \theta) = 1$ and $\lambda = 0$.  At center-of-mass
energies above 195\,GeV,
$C_{\Gamma_{\mathrm{Z}}}(s, \cos \theta)$ is within 5\% of unity.

%
%

In our fit, Equation~(\ref{eq:desh}) is used 
only to compute the expected difference in 
cross section between the Standard Model with and without LSG switched on.
As in the case of the anomalous couplings, the prediction of 
ZZTO for the Standard
Model cross section is used and the expected angular dependence is taken from
YFSZZ.

\begin{figure}
\begin{center}
\epsfig{file=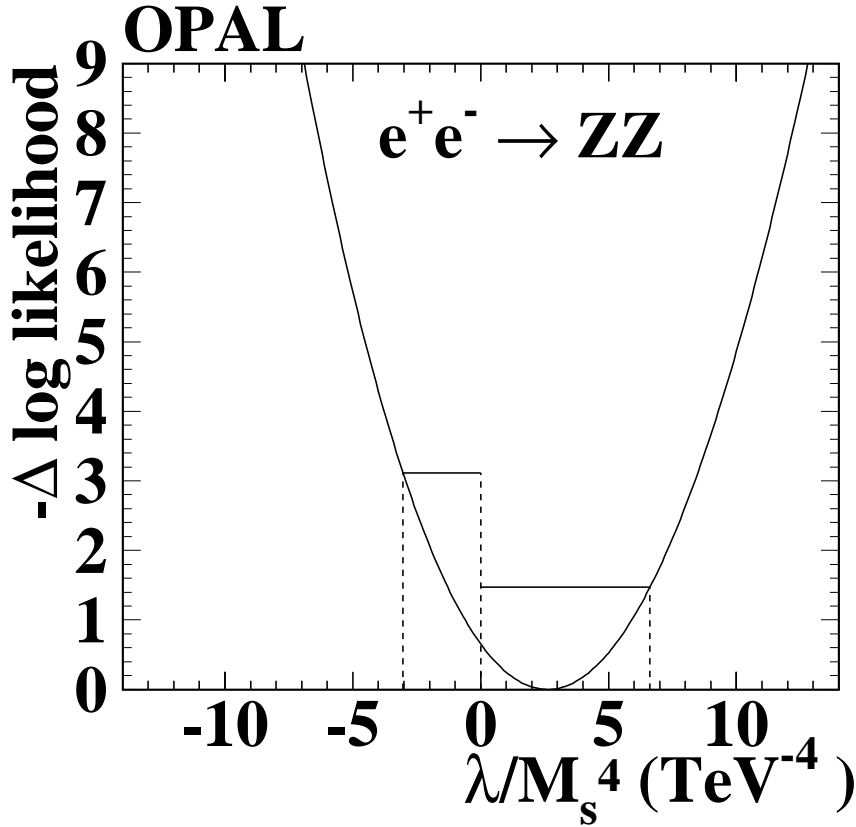,width=0.70\textwidth}
\end{center}
\caption[ED likelihood curve]{
\label{fig:edlike}
The negative log likelihood as a function of the
$\lambda/M^4_s$.  
Limits on $M_s$ are determined separately for theories with
$\lambda = +1$, which correspond to positive
values of $\lambda/M^4_s$   and $\lambda = -1$ which
correspond to negative values of  $\lambda/M^4_s$.
The dashed lines indicate the allowed 95\% confidence level
regions obtained
in the two cases.
These results are dominated by the $190 - 209$\,GeV data, 
but also include the OPAL data from
183\,GeV and 189\,GeV presented in Reference~\cite{bib:zzopal}.
}
\end{figure}

A maximum likelihood fit is performed in 
four bins of $|\cos \theta|$ for each selection and
energy point.
We also use previously published~\cite{bib:zzopal} 
data binned in four
bins of  $|\cos \theta|$ from 189\,GeV 
and the integrated cross section from 183\,GeV.
The fit yields
$$
\lambda/M_s^4 = 2.6 \pm 2.3 \ \mathrm{TeV}^{-4}.
$$   
The likelihood function is shown in
Figure~\ref{fig:edlike} and is approximately parabolic.
Note that with the sign convention of Reference~\cite{bib:joanne}
positive values of $\lambda/M_s^4$ give a positive contribution
to the cross section.  Because our measured values of the 
cross sections
are on average slightly larger than the Standard Model prediction
(see Figure~\ref{fig:zz_cos} and 
the value for $R$ given in Section~\ref{sec:xsec}),
the central value  for $\lambda/M_s^4$ is also positive.
%
%

If we 	assume that the {\em a priori} probability for a theory
to be true is uniform in  the variable $\lambda/M_s^4$ we
can obtain limits on $M_s$ for a variety of theoretical
approaches using the likelihood curve for $\lambda/M_s^4$.
First, using the notation of Reference~\cite{bib:joanne}, we consider
separately theories with
$\lambda = +1$, which correspond to positive
values of $\lambda/M^4_s$, and
with $\lambda = -1$ which correspond to negative values of  $\lambda/M^4_s$.
In the first (second) case  
our  prior assumes all values of  $\lambda/M^4_s > 0.0$ 
( $\lambda/M^4_s < 0.0$) are
equally likely.  We also use the approximation
that the log-likelihood curve is parabolic.
The resulting limits on $M_s$ are given in Table~\ref{tab:edlims}.

In the approach of Reference~\cite{bib:lykken} 
a summation of higher order terms is used which eliminates
the dependence of the prediction on $\lambda$.  
Comparing
References~\cite{bib:joanne} and \cite{bib:lykken}
limits on $M_s$ can be determined,
as a function of the number of large extra dimensions, 
$n$,  using the substitution
\begin{equation}
\frac{\lambda}{M_s^4} \Rightarrow - \pi \frac{s^{\frac{n}{2}-1} }{M_s^{n+2}} 
                               I_n\left( \frac{M_s}{\sqrt{s}} \right)
\end{equation}
where the integrals $I_n ( \frac{M_s}{\sqrt{s}})$ are given in
Equation B.8 of Reference~\cite{bib:lykken}.  As the 
integrals $I_n$ are negative for all values of $n$, limits
on $M_s$ correspond to the $\lambda = +1$  case shown in
Figure~\ref{fig:edlike}.
These limits on $M_s$ as a function of $n$ are
given in Table~\ref{tab:edlims}.


\begin{table}
\begin{center}
\begin{tabular}{|c|c|}
\hline
Parameter      & 95\% C.L. lower limit on $M_s$ \\
\hline
\hline
Coupling        &  Method of Reference~\cite{bib:joanne} \\ 
\hline
$\lambda = +1  $   & 0.62 TeV            \\
$\lambda = -1  $   & 0.76 TeV            \\
\hline
\hline
Number of extra dimensions &   Method of Reference~\cite{bib:lykken}\\
\hline
$n=2$ & 0.92 TeV            \\
$n=3$ & 0.82 TeV            \\
$n=4$ & 0.73 TeV            \\
$n=5$ & 0.67 TeV            \\
$n=6$ & 0.62 TeV            \\
$n=7$ & 0.59 TeV            \\
\hline

\end{tabular}
\end{center}
\caption{
\label{tab:edlims} The 95\% confidence level on $M_s$ using
the approach of References~\cite{bib:joanne} and  \cite{bib:lykken}.
These results include all channels
and all energies,  
including the OPAL data from
183\,GeV and 189\,GeV presented in Reference~\cite{bib:zzopal}.
}
\end{table}

\clearpage
\section{Conclusion} 

The process $\eetozz$ has been studied 
at center-of-mass energies between 190\,GeV and 209\,GeV
using the final states
$\llll$, $\llnn$, $\qqll$, $\qqnn$, and  $\qqqq$.
The number of observed events, the
background expectation from Monte Carlo 
and the calculated efficiencies have been combined to measure
the NC2 cross section for the process $\eetozz$.  
The measured cross sections for the six energy points
are:
$$
\begin{array}{lcll}
\xsec  (192 \ \mathrm{GeV})   & = & \xscc & \mathrm{pb} \\
&&& \\
\xsec  (196 \ \mathrm{GeV})  & = & \xscd & \mathrm{pb} \\
&&& \\
\xsec  (200 \ \mathrm{GeV})   & = & \xsce & \mathrm{pb} \\
&&& \\
\xsec  (202 \ \mathrm{GeV})  & = & \xscf & \mathrm{pb} \\
&&& \\
\xsec  (205 \ \mathrm{GeV})  & = & \xscg & \mathrm{pb} \\
&&& \\
\xsec  (207 \ \mathrm{GeV})  & = & \xsch & \mathrm{pb} .\\
\end{array}
$$
The measurements at all  energies are consistent
with the Standard Model expectations.
Using information from the cross-section measurements
and from the optimal observables, 
no evidence is found for anomalous neutral-current triple gauge
couplings.  The 95\% confidence level limits are listed 
in Table~\ref{tab:anlim}.  We have also derived limits on
low scale gravity theories with large extra dimensions which are summarized
in Table~\ref{tab:edlims}.

\section*{Acknowledgements}

We particularly wish to thank the SL Division for the efficient operation
of the LEP accelerator at all energies
 and for their close cooperation with
our experimental group.  In addition to the support staff at our own
institutions we are pleased to acknowledge the  \\
Department of Energy, USA, \\
National Science Foundation, USA, \\
Particle Physics and Astronomy Research Council, UK, \\
Natural Sciences and Engineering Research Council, Canada, \\
Israel Science Foundation, administered by the Israel
Academy of Science and Humanities, \\
Benoziyo Center for High Energy Physics,\\
Japanese Ministry of Education, Culture, Sports, Science and
Technology (MEXT) and a grant under the MEXT International
Science Research Program,\\
Japanese Society for the Promotion of Science (JSPS),\\
German Israeli Bi-national Science Foundation (GIF), \\
Bundesministerium f\"ur Bildung und Forschung, Germany, \\
National Research Council of Canada, \\
Hungarian Foundation for Scientific Research, OTKA T-038240, 
and T-042864,\\
The NWO/NATO Fund for Scientific Research, the Netherlands.\\

\begin{appendix}
\section*{Appendix~~~Selected events at individual energies}
\label{sec:a}

The information which was used in the 
cross-section fit 
is summarized in Tables~\ref{tab:zzxsec192} to~\ref{tab:zzxsec207}.  
For each channel, the tables give the number of events observed, 
$\nobs$, the Standard Model prediction for
all events, $\smtot$, the expected
signal, $\nsm$, 
the expected background, $\nback$, 
the efficiency $\eff$, 
and the integrated luminosity, $\lint$.  Note that $\smtot$
as it appears in the tables
is not always exactly $\nback + \nsm$ due to rounding.

\begin{table}[b]
\begin{center}
\vskip 0.2cm
$\roots = 192$\,GeV \\
\vskip 0.2cm
{\small
\begin{tabular}{|c|l|ccccccc|}
\hline
&Selection & $\nobs$ & $\smtot$ &$\nsm$ & $\nback$ &$\eff$ & $\br$ & $\lint$\\
&          &         &          &       &          &       &       &
 $(\pb^{-1})$  \\
\hline
 a&$\llllz$&  0&$ 0.28 \pm  0.05$ & $ 0.13 \pm  0.01$ & $ 0.14 \pm  0.05$ & $0.57 \pm 0.02$ &  0.010&  29.1 \\ 
 b&$\eennz$&  0&$ 0.21 \pm  0.03$ & $ 0.12 \pm  0.02$ & $ 0.09 \pm  0.03$ & $0.38 \pm 0.05$ &  0.013&  29.1 \\ 
 c&$\mmnnz$&  0&$ 0.24 \pm  0.04$ & $ 0.13 \pm  0.02$ & $ 0.11 \pm  0.03$ & $0.43 \pm 0.06$ &  0.013&  29.1 \\ 
 d&$\qqeez$&  0&$ 0.65 \pm  0.04$ & $ 0.53 \pm  0.02$ & $ 0.12 \pm  0.03$ & $0.64 \pm 0.03$ &  0.037&  28.6 \\ 
 e&$\bbeez$&  1&$ 0.13 \pm  0.01$ & $ 0.11 \pm  0.01$ & $ 0.02 \pm  0.01$ & $0.47 \pm 0.05$ &  0.010&  28.6 \\ 
 f&$\qqmmz$&  2&$ 0.73 \pm  0.04$ & $ 0.65 \pm  0.02$ & $ 0.08 \pm  0.03$ & $0.79 \pm 0.03$ &  0.037&  28.6 \\ 
 g&$\bbmmz$&  0&$ 0.15 \pm  0.01$ & $ 0.13 \pm  0.01$ & $ 0.01 \pm  0.01$ & $0.59 \pm 0.05$ &  0.010&  28.6 \\ 
 h&$\qqttz$&  0&$ 0.28 \pm  0.03$ & $ 0.25 \pm  0.02$ & $ 0.02 \pm  0.01$ & $0.31 \pm 0.03$ &  0.037&  28.6 \\ 
 i&$\bbttz$&  0&$ 0.07 \pm  0.01$ & $ 0.05 \pm  0.01$ & $ 0.02 \pm  0.01$ & $0.23 \pm 0.04$ &  0.010&  28.6 \\ 
 j&$\xbttz$&  0&$ 0.03 \pm  0.01$ & $ 0.02 \pm  0.01$ & $ 0.01 \pm  0.01$ & $0.10 \pm 0.03$ &  0.010&  28.6 \\ 
 k&$\qqnnz$&  3&$ 3.06 \pm  0.22$ & $ 1.67 \pm  0.14$ & $ 1.40 \pm  0.17$ & $0.34 \pm 0.03$ &  0.219&  28.6 \\ 
 l&$\bbnnz$&  1&$ 0.43 \pm  0.05$ & $ 0.36 \pm  0.03$ & $ 0.07 \pm  0.03$ & $0.27 \pm 0.03$ &  0.061&  28.6 \\ 
 m&$\qqqqz$& 27&$ 17.6 \pm   1.4$ & $  3.8 \pm   0.3$ & $ 13.8 \pm   1.4$ & $0.56 \pm 0.04$ &  0.300&  29.1 \\ 
 n&$\qqxbz$&  2&$ 0.93 \pm  0.11$ & $ 0.52 \pm  0.04$ & $ 0.42 \pm  0.10$ & $0.12 \pm 0.01$ &  0.189&  29.1 \\ 
 o&$\qqbbz$&  1&$ 1.68 \pm  0.13$ & $ 1.21 \pm  0.07$ & $ 0.47 \pm  0.11$ & $0.28 \pm 0.02$ &  0.189&  29.1 \\ 
\hline
\end{tabular}     
}
\end{center}
\caption[192 summary]{ 
The 192\,GeV data.  The description of the entries is
given in the caption of Table~\ref{tab:zzxsum}.
\label{tab:zzxsec192}
}
\end{table}

\begin{table}
\begin{center}
\vskip 0.2cm
$\roots = 196$\,GeV \\
\vskip 0.2cm
{\small
\begin{tabular}{|c|l|ccccccc|}
\hline
&Selection & $\nobs$ & $\smtot$ &$\nsm$ & $\nback$ &$\eff$ & $\br$ & $\lint$\\
&          &         &          &       &          &       &       &
 $(\pb^{-1})$  \\
\hline
 a&$\llllz$&  1&$ 0.58 \pm  0.10$ & $ 0.37 \pm  0.01$ & $ 0.21 \pm  0.10$ & $0.56 \pm 0.02$ &  0.010&  72.5 \\ 
 b&$\eennz$&  1&$ 0.66 \pm  0.08$ & $ 0.37 \pm  0.05$ & $ 0.28 \pm  0.06$ & $0.42 \pm 0.06$ &  0.013&  72.7 \\ 
 c&$\mmnnz$&  0&$ 0.72 \pm  0.10$ & $ 0.37 \pm  0.05$ & $ 0.36 \pm  0.08$ & $0.41 \pm 0.06$ &  0.013&  72.7 \\ 
 d&$\qqeez$&  2&$ 2.05 \pm  0.16$ & $ 1.58 \pm  0.07$ & $ 0.47 \pm  0.14$ & $0.67 \pm 0.03$ &  0.037&  71.3 \\ 
 e&$\bbeez$&  0&$ 0.39 \pm  0.04$ & $ 0.34 \pm  0.03$ & $ 0.05 \pm  0.02$ & $0.52 \pm 0.05$ &  0.010&  71.3 \\ 
 f&$\qqmmz$&  3&$ 1.97 \pm  0.07$ & $ 1.83 \pm  0.06$ & $ 0.14 \pm  0.04$ & $0.77 \pm 0.02$ &  0.037&  71.3 \\ 
 g&$\bbmmz$&  2&$ 0.43 \pm  0.03$ & $ 0.42 \pm  0.02$ & $ 0.01 \pm  0.01$ & $0.63 \pm 0.05$ &  0.010&  71.3 \\ 
 h&$\qqttz$&  0&$ 0.81 \pm  0.07$ & $ 0.68 \pm  0.06$ & $ 0.13 \pm  0.04$ & $0.29 \pm 0.03$ &  0.037&  71.3 \\ 
 i&$\bbttz$&  0&$ 0.19 \pm  0.03$ & $ 0.17 \pm  0.03$ & $ 0.02 \pm  0.02$ & $0.27 \pm 0.04$ &  0.010&  71.3 \\ 
 j&$\xbttz$&  0&$ 0.06 \pm  0.01$ & $ 0.05 \pm  0.01$ & $ 0.01 \pm  0.01$ & $0.08 \pm 0.02$ &  0.010&  71.3 \\ 
 k&$\qqnnz$& 11&$ 8.62 \pm  0.64$ & $ 4.79 \pm  0.43$ & $ 3.83 \pm  0.47$ & $0.34 \pm 0.03$ &  0.219&  70.3 \\ 
 l&$\bbnnz$&  0&$ 1.29 \pm  0.15$ & $ 1.00 \pm  0.10$ & $ 0.30 \pm  0.11$ & $0.26 \pm 0.03$ &  0.061&  70.3 \\ 
 m&$\qqqqz$& 38&$ 42.1 \pm   3.3$ & $ 10.3 \pm   0.7$ & $ 31.8 \pm   3.3$ & $0.50 \pm 0.03$ &  0.300&  75.4 \\ 
 n&$\qqxbz$&  4&$ 3.13 \pm  0.33$ & $ 1.89 \pm  0.14$ & $ 1.24 \pm  0.30$ & $0.15 \pm 0.01$ &  0.189&  75.4 \\ 
 o&$\qqbbz$&  8&$ 4.69 \pm  0.39$ & $ 3.31 \pm  0.21$ & $ 1.38 \pm  0.32$ & $0.26 \pm 0.02$ &  0.189&  75.4 \\ 
\hline
\end{tabular}     
}
\end{center}
\caption[196 summary]{ 
The 196\,GeV data.  The description of the entries is
given in the caption of Table~\ref{tab:zzxsum}.
\label{tab:zzxsec196}
}
\end{table}
\begin{table}
\begin{center}
\vskip 0.2cm
$\roots = 200$\,GeV \\
{\small
\vskip 0.2cm
\begin{tabular}{|c|l|ccccccc|}
\hline
&Selection & $\nobs$ & $\smtot$ &$\nsm$ & $\nback$ &$\eff$ & $\br$ & $\lint$\\
&          &         &          &       &          &       &       &
 $(\pb^{-1})$  \\
\hline
 a&$\llllz$&  0&$ 0.47 \pm  0.09$ & $ 0.43 \pm  0.01$ & $ 0.05 \pm  0.08$ & $0.56 \pm 0.02$ &  0.010&  75.1 \\ 
 b&$\eennz$&  0&$ 0.63 \pm  0.08$ & $ 0.43 \pm  0.06$ & $ 0.21 \pm  0.06$ & $0.43 \pm 0.06$ &  0.013&  74.3 \\ 
 c&$\mmnnz$&  0&$ 0.68 \pm  0.09$ & $ 0.39 \pm  0.06$ & $ 0.28 \pm  0.07$ & $0.40 \pm 0.06$ &  0.013&  74.3 \\ 
 d&$\qqeez$&  3&$ 1.98 \pm  0.12$ & $ 1.69 \pm  0.08$ & $ 0.30 \pm  0.09$ & $0.63 \pm 0.03$ &  0.037&  73.7 \\ 
 e&$\bbeez$&  0&$ 0.46 \pm  0.04$ & $ 0.41 \pm  0.03$ & $ 0.04 \pm  0.03$ & $0.56 \pm 0.05$ &  0.010&  73.7 \\ 
 f&$\qqmmz$&  5&$ 2.19 \pm  0.08$ & $ 2.01 \pm  0.06$ & $ 0.18 \pm  0.05$ & $0.75 \pm 0.02$ &  0.037&  73.7 \\ 
 g&$\bbmmz$&  3&$ 0.43 \pm  0.03$ & $ 0.41 \pm  0.03$ & $ 0.02 \pm  0.02$ & $0.56 \pm 0.05$ &  0.010&  73.7 \\ 
 h&$\qqttz$&  1&$ 0.90 \pm  0.08$ & $ 0.76 \pm  0.07$ & $ 0.15 \pm  0.05$ & $0.28 \pm 0.03$ &  0.037&  73.7 \\ 
 i&$\bbttz$&  0&$ 0.23 \pm  0.03$ & $ 0.21 \pm  0.03$ & $ 0.02 \pm  0.02$ & $0.28 \pm 0.04$ &  0.010&  73.7 \\ 
 j&$\xbttz$&  0&$ 0.07 \pm  0.02$ & $ 0.06 \pm  0.02$ & $ 0.02 \pm  0.02$ & $0.07 \pm 0.02$ &  0.010&  73.7 \\ 
 k&$\qqnnz$& 11&$ 10.2 \pm   0.7$ & $  5.4 \pm   0.5$ & $ 4.79 \pm  0.46$ & $0.34 \pm 0.03$ &  0.219&  73.4 \\ 
 l&$\bbnnz$&  0&$ 1.46 \pm  0.15$ & $ 1.23 \pm  0.13$ & $ 0.23 \pm  0.07$ & $0.28 \pm 0.03$ &  0.061&  73.4 \\ 
 m&$\qqqqz$& 26&$ 32.4 \pm   2.5$ & $  9.1 \pm   0.6$ & $ 23.3 \pm   2.4$ & $0.40 \pm 0.03$ &  0.300&  77.7 \\ 
 n&$\qqxbz$&  6&$ 4.33 \pm  0.41$ & $ 2.73 \pm  0.18$ & $ 1.60 \pm  0.37$ & $0.19 \pm 0.02$ &  0.189&  77.7 \\ 
 o&$\qqbbz$&  2&$ 4.26 \pm  0.32$ & $ 3.16 \pm  0.19$ & $ 1.10 \pm  0.26$ & $0.22 \pm 0.02$ &  0.189&  77.7 \\ 
\hline
\end{tabular}     
}
\end{center}
\caption[200 summary]{ 
The 200\,GeV data.  The description of the entries is
given in the caption of Table~\ref{tab:zzxsum}.
\label{tab:zzxsec200}
}
\end{table}
\begin{table}
\begin{center}
\vskip 0.2cm
$\roots = 202$\,GeV \\
\vskip 0.2cm
{\small
\begin{tabular}{|c|l|ccccccc|}
\hline
&Selection & $\nobs$ & $\smtot$ &$\nsm$ & $\nback$ &$\eff$ & $\br$ & $\lint$\\
&          &         &          &       &          &       &       &
 $(\pb^{-1})$  \\
\hline
 a&$\llllz$&  0&$ 0.27 \pm  0.05$ & $ 0.22 \pm  0.01$ & $ 0.04 \pm  0.05$ & $0.56 \pm 0.02$ &  0.010&  38.3 \\ 
 b&$\eennz$&  0&$ 0.37 \pm  0.04$ & $ 0.23 \pm  0.03$ & $ 0.13 \pm  0.03$ & $0.46 \pm 0.06$ &  0.013&  37.1 \\ 
 c&$\mmnnz$&  0&$ 0.44 \pm  0.05$ & $ 0.23 \pm  0.03$ & $ 0.22 \pm  0.04$ & $0.44 \pm 0.06$ &  0.013&  37.1 \\ 
 d&$\qqeez$&  2&$ 0.99 \pm  0.06$ & $ 0.85 \pm  0.04$ & $ 0.14 \pm  0.05$ & $0.62 \pm 0.03$ &  0.037&  36.6 \\ 
 e&$\bbeez$&  1&$ 0.19 \pm  0.02$ & $ 0.17 \pm  0.02$ & $ 0.02 \pm  0.01$ & $0.46 \pm 0.05$ &  0.010&  36.6 \\ 
 f&$\qqmmz$&  0&$ 1.10 \pm  0.04$ & $ 1.01 \pm  0.03$ & $ 0.09 \pm  0.03$ & $0.73 \pm 0.02$ &  0.037&  36.6 \\ 
 g&$\bbmmz$&  0&$ 0.25 \pm  0.02$ & $ 0.24 \pm  0.01$ & $ 0.01 \pm  0.01$ & $0.64 \pm 0.05$ &  0.010&  36.6 \\ 
 h&$\qqttz$&  1&$ 0.47 \pm  0.04$ & $ 0.41 \pm  0.04$ & $ 0.06 \pm  0.02$ & $0.30 \pm 0.03$ &  0.037&  36.6 \\ 
 i&$\bbttz$&  0&$ 0.12 \pm  0.02$ & $ 0.11 \pm  0.02$ & $ 0.01 \pm  0.01$ & $0.28 \pm 0.04$ &  0.010&  36.6 \\ 
 j&$\xbttz$&  0&$ 0.04 \pm  0.01$ & $ 0.03 \pm  0.01$ & $ 0.01 \pm  0.01$ & $0.07 \pm 0.02$ &  0.010&  36.6 \\ 
 k&$\qqnnz$&  3&$ 4.66 \pm  0.32$ & $ 2.53 \pm  0.23$ & $ 2.13 \pm  0.23$ & $0.31 \pm 0.03$ &  0.219&  36.3 \\ 
 l&$\bbnnz$&  1&$ 0.71 \pm  0.07$ & $ 0.64 \pm  0.07$ & $ 0.08 \pm  0.03$ & $0.28 \pm 0.03$ &  0.061&  36.3 \\ 
 m&$\qqqqz$& 22&$ 21.5 \pm   1.7$ & $  5.4 \pm   0.4$ & $ 16.1 \pm   1.7$ & $0.48 \pm 0.03$ &  0.300&  36.6 \\ 
 n&$\qqxbz$&  2&$ 1.54 \pm  0.14$ & $ 1.03 \pm  0.08$ & $ 0.51 \pm  0.12$ & $0.15 \pm 0.01$ &  0.189&  36.6 \\ 
 o&$\qqbbz$&  0&$ 2.21 \pm  0.17$ & $ 1.63 \pm  0.10$ & $ 0.59 \pm  0.14$ & $0.23 \pm 0.02$ &  0.189&  36.6 \\ 
\hline
\end{tabular}     
}
\end{center}
\caption[202 summary]{ 
The 202\,GeV data.  The description of the entries is
given in the caption of Table~\ref{tab:zzxsum}.
\label{tab:zzxsec202}
}
\end{table}

\begin{table}
\begin{center}
\vskip 0.2cm
$\roots = 205$\,GeV \\
\vskip 0.2cm
{\small
\begin{tabular}{|c|l|ccccccc|}
\hline
&Selection & $\nobs$ & $\smtot$ &$\nsm$ & $\nback$ &$\eff$ & $\br$ & $\lint$\\
&          &         &          &       &          &       &       &
 $(\pb^{-1})$  \\
\hline
 a&$\llllz$&  1&$ 0.73 \pm  0.12$ & $ 0.49 \pm  0.01$ & $ 0.24 \pm  0.12$ & $0.55 \pm 0.02$ &  0.010&  81.7 \\ 
 b&$\eennz$&  0&$ 0.68 \pm  0.09$ & $ 0.47 \pm  0.06$ & $ 0.21 \pm  0.06$ & $0.39 \pm 0.05$ &  0.013&  84.4 \\ 
 c&$\mmnnz$&  0&$ 0.81 \pm  0.11$ & $ 0.48 \pm  0.07$ & $ 0.33 \pm  0.08$ & $0.40 \pm 0.06$ &  0.013&  84.4 \\ 
 d&$\qqeez$&  1&$ 2.19 \pm  0.12$ & $ 1.83 \pm  0.09$ & $ 0.36 \pm  0.08$ & $0.59 \pm 0.03$ &  0.037&  80.5 \\ 
 e&$\bbeez$&  1&$ 0.51 \pm  0.05$ & $ 0.44 \pm  0.04$ & $ 0.06 \pm  0.03$ & $0.51 \pm 0.05$ &  0.010&  80.5 \\ 
 f&$\qqmmz$&  4&$ 2.56 \pm  0.09$ & $ 2.43 \pm  0.08$ & $ 0.13 \pm  0.05$ & $0.78 \pm 0.03$ &  0.037&  80.5 \\ 
 g&$\bbmmz$&  2&$ 0.51 \pm  0.04$ & $ 0.47 \pm  0.03$ & $ 0.04 \pm  0.03$ & $0.55 \pm 0.05$ &  0.010&  80.5 \\ 
 h&$\qqttz$&  0&$ 1.06 \pm  0.10$ & $ 0.92 \pm  0.08$ & $ 0.13 \pm  0.05$ & $0.30 \pm 0.03$ &  0.037&  80.5 \\ 
 i&$\bbttz$&  0&$ 0.31 \pm  0.04$ & $ 0.25 \pm  0.04$ & $ 0.06 \pm  0.02$ & $0.29 \pm 0.05$ &  0.010&  80.5 \\ 
 j&$\xbttz$&  1&$ 0.06 \pm  0.02$ & $ 0.05 \pm  0.02$ & $ 0.01 \pm  0.01$ & $0.06 \pm 0.02$ &  0.010&  80.5 \\ 
 k&$\qqnnz$&  7&$ 10.9 \pm   0.7$ & $  5.8 \pm   0.5$ & $ 5.10 \pm  0.50$ & $0.31 \pm 0.03$ &  0.219&  79.9 \\ 
 l&$\bbnnz$&  0&$ 1.68 \pm  0.18$ & $ 1.48 \pm  0.15$ & $ 0.20 \pm  0.09$ & $0.29 \pm 0.03$ &  0.061&  79.9 \\ 
 m&$\qqqqz$& 36&$ 25.1 \pm   1.9$ & $  8.0 \pm   0.6$ & $ 17.0 \pm   1.8$ & $0.32 \pm 0.02$ &  0.300&  80.2 \\ 
 n&$\qqxbz$&  3&$ 4.46 \pm  0.42$ & $ 2.82 \pm  0.18$ & $ 1.64 \pm  0.38$ & $0.18 \pm 0.01$ &  0.189&  80.2 \\ 
 o&$\qqbbz$&  4&$ 3.18 \pm  0.24$ & $ 2.51 \pm  0.17$ & $ 0.67 \pm  0.18$ & $0.16 \pm 0.01$ &  0.189&  80.2 \\ 
\hline
\end{tabular}     
}
\end{center}
\caption[205 summary]{ 
The 205\,GeV data.  The description of the entries is
given in the caption of Table~\ref{tab:zzxsum}.
\label{tab:zzxsec205}
}
\end{table}

\begin{table}
\begin{center}
\vskip 0.2cm
$\roots = 207$\,GeV \\
\vskip 0.2cm
{\small
\begin{tabular}{|c|l|ccccccc|}
\hline
&Selection & $\nobs$ & $\smtot$ &$\nsm$ & $\nback$ &$\eff$ & $\br$ & $\lint$\\
&          &         &          &       &          &       &       &
 $(\pb^{-1})$  \\
\hline
 a&$\llllz$&  2&$ 1.23 \pm  0.20$ & $ 0.83 \pm  0.03$ & $ 0.40 \pm  0.20$ & $0.55 \pm 0.02$ &  0.010& 136.9 \\ 
 b&$\eennz$&  1&$ 1.16 \pm  0.14$ & $ 0.81 \pm  0.11$ & $ 0.36 \pm  0.10$ & $0.41 \pm 0.05$ &  0.013& 137.6 \\ 
 c&$\mmnnz$&  0&$ 1.41 \pm  0.19$ & $ 0.79 \pm  0.11$ & $ 0.62 \pm  0.15$ & $0.40 \pm 0.06$ &  0.013& 137.6 \\ 
 d&$\qqeez$&  6&$ 3.72 \pm  0.21$ & $ 3.12 \pm  0.16$ & $ 0.59 \pm  0.13$ & $0.59 \pm 0.03$ &  0.037& 134.0 \\ 
 e&$\bbeez$&  0&$ 0.82 \pm  0.08$ & $ 0.72 \pm  0.07$ & $ 0.10 \pm  0.05$ & $0.49 \pm 0.05$ &  0.010& 134.0 \\ 
 f&$\qqmmz$&  1&$ 4.37 \pm  0.15$ & $ 4.16 \pm  0.13$ & $ 0.21 \pm  0.08$ & $0.79 \pm 0.03$ &  0.037& 134.0 \\ 
 g&$\bbmmz$&  0&$ 0.82 \pm  0.07$ & $ 0.75 \pm  0.05$ & $ 0.07 \pm  0.04$ & $0.51 \pm 0.05$ &  0.010& 134.0 \\ 
 h&$\qqttz$&  2&$ 1.83 \pm  0.16$ & $ 1.61 \pm  0.14$ & $ 0.22 \pm  0.08$ & $0.30 \pm 0.03$ &  0.037& 134.0 \\ 
 i&$\bbttz$&  0&$ 0.49 \pm  0.08$ & $ 0.40 \pm  0.06$ & $ 0.09 \pm  0.04$ & $0.27 \pm 0.05$ &  0.010& 134.0 \\ 
 j&$\xbttz$&  0&$ 0.09 \pm  0.03$ & $ 0.08 \pm  0.03$ & $ 0.01 \pm  0.01$ & $0.06 \pm 0.02$ &  0.010& 134.0 \\ 
 k&$\qqnnz$& 16&$ 18.9 \pm   1.2$ & $ 10.1 \pm   0.9$ & $ 8.80 \pm  0.86$ & $0.32 \pm 0.03$ &  0.219& 133.6 \\ 
 l&$\bbnnz$&  7&$ 2.88 \pm  0.30$ & $ 2.53 \pm  0.26$ & $ 0.35 \pm  0.15$ & $0.29 \pm 0.03$ &  0.061& 133.6 \\ 
 m&$\qqqqz$& 36&$ 41.8 \pm   3.1$ & $ 13.6 \pm   0.9$ & $ 28.3 \pm   3.0$ & $0.32 \pm 0.02$ &  0.300& 133.3 \\ 
 n&$\qqxbz$&  7&$ 7.48 \pm  0.69$ & $ 4.75 \pm  0.30$ & $ 2.73 \pm  0.63$ & $0.18 \pm 0.01$ &  0.189& 133.3 \\ 
 o&$\qqbbz$&  6&$ 5.34 \pm  0.41$ & $ 4.23 \pm  0.28$ & $ 1.11 \pm  0.29$ & $0.16 \pm 0.01$ &  0.189& 133.3 \\ 

\hline
\end{tabular}    
} 
\end{center}
\caption[207 summary]{ 
The 207\,GeV data.  The description of the entries is
given in the caption of Table~\ref{tab:zzxsum}.
\label{tab:zzxsec207}
}
\end{table}
\end{appendix}
\clearpage 

\end{document}